%% file: ilcanomalies.tex
\documentclass[a4paper,11pt]{article}
\pdfoutput=1 

\usepackage{jheppub} 
\usepackage{ilcanomalies-defs}
\usepackage[T1]{fontenc} 
\usepackage{tikz}
\usepackage[compat=1.1.0]{tikz-feynman}
\usepackage{caption}
\usepackage{subcaption}
\usepackage{multirow}

\title{\boldmath High-dimensional Anomaly Detection with Radiative Return in \ee~Collisions}

\author[a]{Julia Gonski,}
\author[b]{Jerry Lai,}
\author[c,d]{Benjamin Nachman,}
\author[e]{and In\^{e}s Ochoa}

\affiliation[a]{Nevis Laboratories, Columbia University, 136 S Broadway, Irvington NY, USA}
\affiliation[c]{Department of Electrical Engineering and Computer Sciences, University of California, Berkeley, CA 94720, USA}
\affiliation[c]{Physics Division, Lawrence Berkeley National Laboratory, Berkeley, CA 94720, USA}
\affiliation[d]{Berkeley Institute for Data Science, University of California, Berkeley, CA 94720, USA}
\affiliation[e]{Laboratory of Instrumentation and Experimental Particle Physics, Lisbon, Portugal}

\emailAdd{julia.gonski@cern.ch}
\emailAdd{thejerrylai@berkeley.edu}
\emailAdd{bpnachman@lbl.gov}
\emailAdd{ines.ochoa@cern.ch}

\abstract{Experiments at a future \ee~collider will be able to search for new particles with masses below the nominal centre-of-mass energy by analyzing collisions with initial-state radiation (radiative return).  We show that machine learning methods that use imperfect or missing training labels can achieve sensitivity to generic new particle production in radiative return events.  In addition to presenting an application of the classification without labels (CWoLa) search method in \ee~collisions, our study combines weak supervision with variable-dimensional information by deploying a deep sets neural network architecture.  We have also investigated some of the experimental aspects of anomaly detection in radiative return events and discuss these in the context of future detector design.
}

\begin{document} 
\maketitle
\flushbottom

\input{sections/introduction}
\input{sections/samples}
\input{sections/training}
\input{sections/results}

\input{sections/futuredetector}
\input{sections/conclusion}


\section*{Code and Data}

Our code can be found at \url{https://github.com/bnachman/ILCAnomalies} and the data are available at Zenodo\footnote{\url{https://zenodo.org/record/5181211}}.

\acknowledgments

The authors would like to thank Michael Peskin, for discussions and insight into simulation and analysis of \ee~collisions. 

JG is supported by the National Science Foundation under Grant No. PHY-2013070.  BN is supported by the Department of Energy, Office of Science under contract number DE-AC02-05CH11231. IO is supported by the fellowship LCF/BQ/PI20/11760025 from La Caixa Foundation (ID 100010434) and by the European Union Horizon 2020 research and innovation program under the Marie Sk\l{}odowska-Curie grant agreement No 847648.
\clearpage

\appendix
\appendix
\input{sections/app_pfnvars}
\input{sections/app_eventlevelvars}

\bibliographystyle{jhep}
\bibliography{HEPML,ilcanomalies}

\end{document}

%% file: sections/introduction.tex
\section{Introduction}
\label{sec:intro}

Searches for resonant new physics has been a cornerstone of high energy physics since at least the discovery of the $\rho$ meson~\cite{PhysRev.126.1858} and up to and including the Higgs boson discovery by the ATLAS and CMS Collaborations at the Large Hadron Collider (LHC)~\cite{atlas_higgs,cms_higgs}. Since that time, there have been many proposals to widen the sensitivity of bump hunt analyses using machine learning (ML)~\cite{DAgnolo:2018cun,Collins:2018epr,Collins:2019jip,DAgnolo:2019vbw,Farina:2018fyg,Heimel:2018mkt,Roy:2019jae,Cerri:2018anq,Blance:2019ibf,Hajer:2018kqm,DeSimone:2018efk,1809.02977,Dillon:2019cqt,Andreassen:2020nkr,Nachman:2020lpy,Aguilar-Saavedra:2017rzt,Romao:2019dvs,Romao:2020ojy,knapp2020adversarially,collaboration2020dijet,1797846,1800445,Amram:2020ykb,Cheng:2020dal,Khosa:2020qrz,aguilarsaavedra2020mass,1815227,pol2020anomaly,Mikuni:2020qds,vanBeekveld:2020txa,Park:2020pak,Faroughy:2020gas,Stein:2020rou,Kasieczka:2021xcg,Blance:2021gcs,Bortolato:2021zic,Collins:2021nxn,Dillon:2021nxw,Finke:2021sdf,Atkinson:2021nlt,Kahn:2021drv,Aarrestad:2021oeb,Dorigo:2021iyy,Caron:2021wmq,Govorkova:2021hqu,Kasieczka:2021tew} (see also Ref.~\cite{Feickert:2021ajf}).  While searching for a peak in the invariant mass of two or more objects is a generic strategy for searching for the signature of new particles, it is not particularly sensitive to any given new physics model.  Recent ML proposals identify patterns in high dimensional data, which can improve discovery potential while maintaining a low degree of signal-model and background-model dependence.  The ML-assisted bump hunt has also been extended beyond the LHC to astrophysics, in the pursuit of stellar streams~\cite{Shih:2021kbt}.  

The goal of the current paper is to explore the potential of resonant anomaly detection at a future $e^+e^-$ collider. The identification of anomalous elements is done through a particular class of algorithms based on weakly supervised learning.  Most ML applications in high energy physics are based on supervised learning, whereby training examples are produced from simulations with a known origin.  Training ML methods directly on data must be able to cope with the lack of labels for what inputs are background or a potential new signal. The strategy used here leverages noisy labels on high-dimensional hadronic final state inputs. 

The choice of studying advanced analysis strategies at future colliders is timely given the ongoing discussions of next generation energy frontier experiments~\cite{FCC:2018byv,FCC:2018evy,FCC:2018vvp,Behnke:2013xla,Baer:2013cma,Adolphsen:2013jya,Adolphsen:2013kya,Behnke:2013lya,InternationalLinearColliderInternationalDevelopmentTeam:2021guz,CEPCStudyGroup:2018rmc,CEPCStudyGroup:2018ghi,Linssen:1425915,CLICdp:2018cto}.  This is particularly true if new techniques require dedicated detector modifications or computing resources.  Additionally, an increased sensitivity from ML methods may increase the case for one future machine over another. This result explores the use of machine learning without signal/background labels (weak supervision) in $e^+e^-$ collisions and the input features can be high-dimensional and vary in size from event to event (variable-dimensional\footnote{Earlier studies of weak supervision with high-dimensional features can be found in Ref.~\cite{Komiske:2018oaa} and earlier studies of high-dimensional weak supervision for anomaly detection can be found in Ref.~\cite{Amram:2020ykb}.}).


Unlike at a hadron collider, the particles in a lepton collider are designed to have a fixed center-of-mass energy ($\sqrt{s}$). One can perform resonant anomaly detection by sifting through the collision debris for a range of center-of-mass energies.  While beam energy scans are one approach to exploring different $\sqrt{s}$~values, we consider a complementary setting in this paper: \textit{radiative return}.  Due to initial state photon radiation that can carry a variable amount of energy away from the collision, collisions at fixed beam energies span a wide range in $e^+e^-$ $\sqrt{\hat{s}}\leq \sqrt{s}$, where \shat~is the effective collision energy.  These radiative return events are often considered a nuisance for physics analysis, but they may provide a unique setting for new physics searches.  The radiative return final state has been studied extensively for hadron spectroscopy at past $e^+e^-$ colliders~\cite{Denig:2006kj,Kluge:2008fb,Czyz:2010hj,Druzhinin:2011qd} and was considered in Ref.~\cite{Karliner:2015tga} for classical resonance searches at future colliders. We extend this study to the case of an ML-assisted bump hunt.

Machine learning has been studied in the context of future $e^+e^-$ colliders~\cite{Li:2019ufu,Li:2020vav} for probing Higgs boson properties.  In addition to exploring the use of ML for direct new physics searches at a future $e^+e^-$ collider, we also use a deep neural network architecture that is capable of handling complex hadronic final states.  In particular, high energy $e^+e^-$ events can produce a variable number of hadrons, with no inherent order.  This is a point cloud, which can be processed with graph neural networks or set-based architectures.  We use the latter, employing the Deep Sets architecture~\cite{10.5555/3294996.3295098} implemented as a Particle Flow Network (PFN)~\cite{Komiske:2018cqr}.  Most of the previous weakly supervised methods used at most a handful of high-level observables, while we will be able to use high-dimensional, low-level inputs.

This paper is organized as follows. The simulations used for our empirical studies are presented in Sec.~\ref{sec:samples}. Section~\ref{sec:training} introduces machine learning-based resonant anomaly detection methodology and technical details related to the training setup.  Our numerical results are shown in Sec.~\ref{sec:results} and a discussion about the impact on future detector design is provided in Sec.~\ref{sec:futuredetector}.  The paper ends with conclusions and outlook in Sec.~\ref{sec:conclusion}.





%% file: sections/samples.tex
\section{Simulated Samples and Processing} \label{sec:samples}

We consider \ee~collisions at a nominal center-of-mass (CoM) energy of $\sqrt{s}=1$~TeV that produce final states with jets and a photon from initial state radiation (ISR). The signal process studied is the production of a BSM heavy scalar $X$ that decays into a pair of scalars $a$, each decaying to two $b$-quarks, in association with an ISR photon: ${e^{+}e^{-} \rightarrow X\gamma \rightarrow aa\gamma \rightarrow b\bar{b}b\bar{b} \gamma}$. The following two sets of values for the invariant masses of particles $X$ and $a$ are examined: ${m_X, m_a = 350, 40}$~GeV and $700, 100$~GeV. The background originates from multijet production in association with an ISR photon, with a cross-section that is dominated by the Drell-Yan $\gamma^{*}/Z$ production and extends to close to the nominal 1~TeV CoM energy. Leading-order Feynman diagrams of the signal and background processes are shown in Fig.~\ref{fig:feyn}.  Note that while we have selected two specific signal models, the methods studied here are largely\footnote{The signal model only plays a role in the choice of the features used for learning and if it does not have a narrow width, in the size of the invariant mass windows.} signal model agnostic.

\input{figures/feynman.tex}

The generation of background and signal events is done with \textsc{MadGraph5\_aMC@NLO} 2.8.0~\cite{Alwall:2014hca}.  The background is generated at next-to-leading order in the strong coupling constant while the signal is simulated at leading order (\textsc{MadGraph} syntax: \verb#e+ e- > j j a [QCD]# and \verb#e+ e- > a h#, respectively, where \verb#a# is the photon).  Additional background processes are not included, as they are small, non-resonant in the regions we target, and thus do not change our conclusions.  Future work will investigate the impact of more heterogeneous background data.  In our study, simulated events are then passed to parton showering and hadronization using \textsc{Pythia}~8.244~\cite{Sjostrand:2014zea} with its default settings. A minimum \Et~threshold of 10~GeV is placed on the photon, with a pseudo-rapidity that extends to $\pm 5$. 

The detector simulation is parameterized with \textsc{Delphes} 3.4.2~\cite{Selvaggi:2014mya}, using a card for a generic ILC detector\footnote{The code for this detector card is available at \url{https://github.com/iLCSoft/ILCDelphes}.  The parameterizations are mostly based on the ILD detector design~\cite{ILDConceptGroup:2020sfq}.}. A particle flow algorithm is used to combine tracking and calorimeter information and define the final reconstructed objects. Photons are built from energy deposits in the electromagnetic calorimeter, using the central and forward calorimeter systems with pseudo-rapidity coverages of $|\eta|<3.0$ and $3.0<|\eta|<4.0$, respectively. Jets are built from particle flow objects (except isolated muons, electrons and photons) measured in the tracker (with an acceptance of up to $|\eta|<3.0$), electromagnetic and hadronic calorimeters (the latter with an acceptance of up to 2.8 in the central system and up to 3.8 in the forward system in absolute pseudo-rapidity). The jet clustering is performed with the anti-$k_{t}$~\cite{antikt} algorithm with a radius $R=1.0$ implemented in \textsc{FastJet}~3.3.2~\cite{Cacciari:2005hq,fastjet}. 

\begin{figure}
\begin{centering}
\includegraphics[scale=0.6]{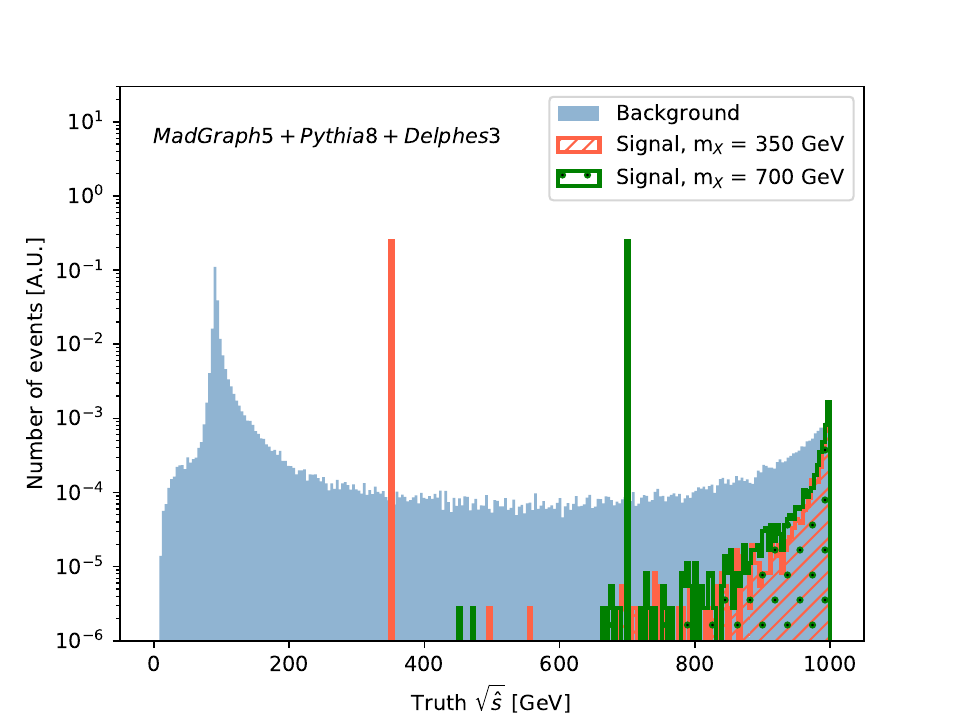}
\caption{Effective center-of-mass energy calculated from truth-level quantities for the background and two signal processes considered. \label{fig:truthsqrtshat}}
\end{centering}
\end{figure}

\begin{figure}
\centering
\includegraphics[scale=0.6]{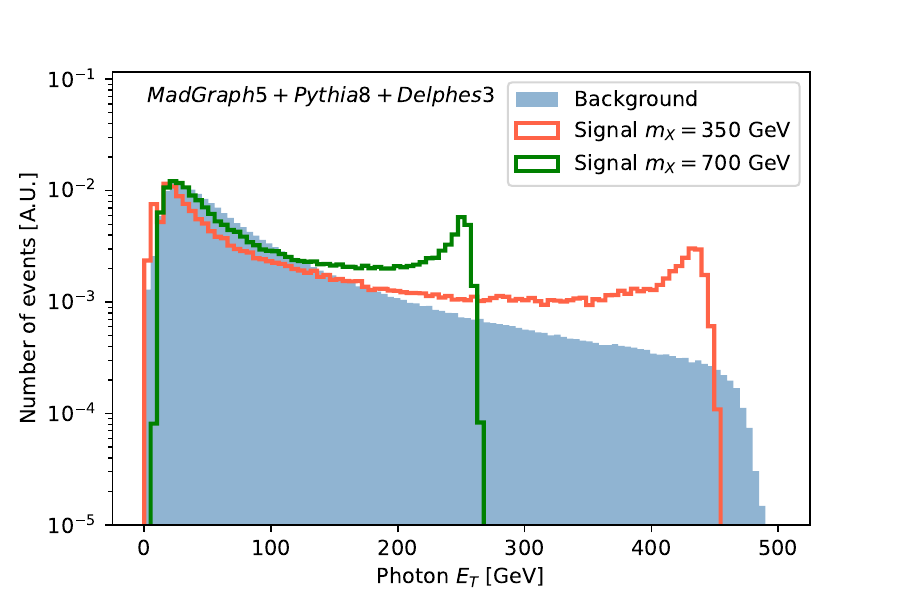}
\includegraphics[scale=0.6]{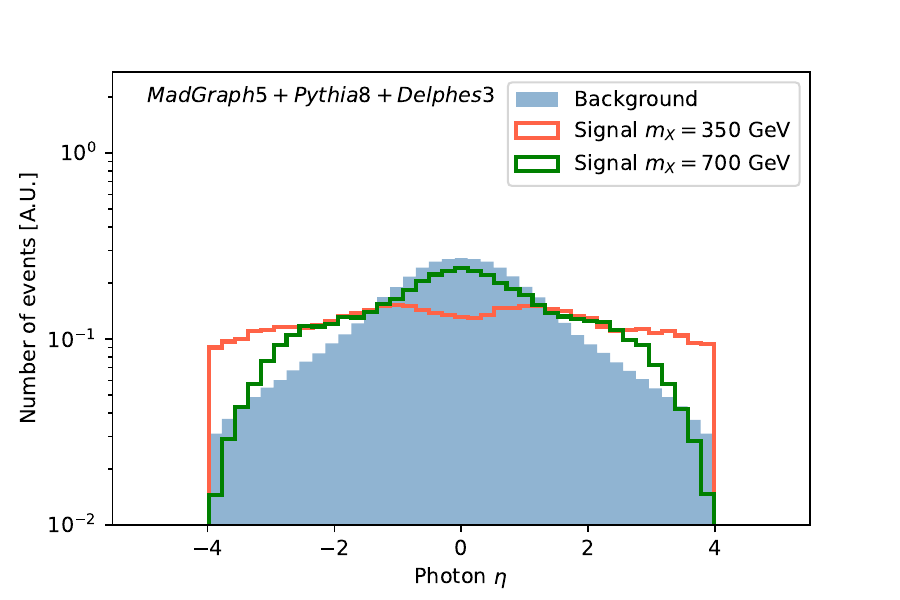}
\caption{Distributions at detector-level of the photon transverse energy and pseudo-rapidity for the background and two signal processes. A minimum cut of 200~GeV is placed on the truth \shat. \label{fig:kin}}
\end{figure}

Events are selected for analysis if they contain at least two jets with a minimum transverse momentum (\pt) of 5~GeV. An effective CoM energy can be calculated for all events based on the energy of the colliding \ee~pair after the photon radiation, as shown in Fig.~\ref{fig:truthsqrtshat} for all generated samples and based on truth-level quantities. Distributions of the photon transverse energy and pseudo-rapidity at detector-level are shown on Fig.~\ref{fig:kin} for the background and signal processes considered.  For detector-level \shat~ studies, we select the highest $p_T$ photon.

%% file: figures/feynman.tex
\begin{figure}
\begin{centering}
\begin{subfigure}[b]{0.3\textwidth}
    \begin{tikzpicture}
      \begin{feynman}
        \vertex (a);
        \vertex [above left=1.3cm of a](i1) {\(e^{-}\)};
        \vertex [below left=1.3cm of a] (i2) {\(e^{+}\)};
        \vertex [below left=0.4cm of a] (i3);
        \vertex [below right=1cm of i3] (c) {\(\gamma\)};
        \vertex [right=1cm of a] (b);
        \vertex [below right=1.3cm of b] (f1) {\(\bar{q}\)};
        \vertex [above right=1.3cm of b] (f2) {\(q\)};
        
        \diagram* {
          (i1)  -- [fermion] (a) -- [fermion] (i2) ,
          (i2) -- [draw=none] (i3) -- [photon] (c),
          (a) -- [photon, edge label=\(\gamma\\ \text{*}/ Z\)] (b),
          (f1) -- [fermion] (b) -- [fermion] (f2),
        };
      \end{feynman}
    \end{tikzpicture}
    \caption{}
\end{subfigure}
\hspace{0.1cm}
\begin{subfigure}[b]{0.3\textwidth}
    \begin{tikzpicture}
      \begin{feynman}
        \vertex (a);
        \vertex [above left=1.3cm of a](i1) {\(e^{-}\)};
        \vertex [below left=1.3cm of a] (i2) {\(e^{+}\)};
        \vertex [below left=0.4cm of a] (i3);
        \vertex [below right=1cm of i3] (c) {\(\gamma\)};
        \vertex [right=1cm of a] (b);
        \vertex [below right=1.3cm of b] (f1) ;
        \vertex [above right=1.3cm of b] (f2) ;
        \vertex [above right=0.5cm of f1] (bb1) {\(\bar{b}\)};
        \vertex [below right=0.5cm of f1] (bb2) {\(b\)};
        \vertex [above right=0.5cm of f2] (bb3) {\(\bar{b}\)};
        \vertex [below right=0.5cm of f2] (bb4) {\(b\)};
        
        \diagram* {
          (i1)  -- [fermion] (a) -- [fermion] (i2) ,
          (i2) -- [draw=none] (i3) -- [photon] (c),
          (a) -- [boson, edge label=\(X\)] (b),
          (f1) -- [scalar, edge label=\(a\)] (b) -- [scalar, edge label=\(\bar{a}\)] (f2),
          (bb1) -- [fermion] (f1) -- [fermion] (bb2),
          (bb3) -- [fermion] (f2) -- [fermion] (bb4),
        };
      \end{feynman}
    \end{tikzpicture}
    \caption{}
\end{subfigure}
\caption{Feynman diagrams of the background (a) and signal (b) processes considered. \label{fig:feyn}}
\end{centering}
\end{figure}

%% file: sections/training.tex
\section{Methodology}
\label{sec:training}

A ML-assisted bump hunt is performed by scanning three different \shat~measures in \ee~collisions: truth \shat, where the truth-level energy of the photon is subtracted from the incoming electron and positron beam energies; the $\gamma$-measured \shat, relying on the measured energy of the detected photon; and the hadron-measured \shat, computed using all the measured particles (in this case hadrons) with the exception of the photon.  The truth-level analysis is presented as nominal, representing the ideal performance of the method without consideration of detector effects in the construction of the signal regions (detector effects are always included for the classification). Further description of the measured \shat~quantities and their impact on performance is given in Section~\ref{sec:futuredetector}.

Signal regions are defined for both mass points as windows of 50 GeV centered at the resonance mass \mx. The sideband region then extends 50 GeV in both directions, excluding the signal region. A summary of these definitions can be found in Table~\ref{tab:regions}. The training (which is described later in more detail) is performed with a fixed luminosity over events from each of the three \shat~measures. The luminosity is determined from a normalization chosen to give 25000 events in the $m_X = 350$~GeV sideband region. A total of 5000225 background and 90001 signal events were generated.

In particular, the unpolarized cross-section for the background process in our fiducial region is about 1~pb (as reported by \textsc{MadGraph5\_aMC@NLO}). This amount of statistics would correspond to a collected luminosity of approximately 6.5~ab\textsuperscript{-1}, well within the integrated luminosity estimates of the operating scenarios for colliders at 1~TeV~\cite{Barklow:2015tja}.  While we have used this particular setup to provide concrete conclusions, the methodology and qualitative conclusions apply more generally to any future \ee~collider.



For a given \shat~distribution, training utilizes one label for background events and another label for pseudodata, which is composed of background and a number of signal contamination events. The number of injected signal events is calculated in terms of the local signal significance $\sigma$ in the signal region, where $\sigma \equiv S/\sqrt{B}$, the approximate statistical significance, where $S$ is the number of signal events and $B$ is the number of background events, both in the pseudodata input class which draws events from the signal region \shat~window. The training for a given configuration is run seven times, with signal contaminations that correspond to $\sigma = 0, 0.5, 1, 2, 3, 5$ and $\infty$ (100\% signal). The background events are the same in each training, while the signal events added to the pseudodata are randomly selected. Event yields used in training for the background and signal samples in each region can be found in Table~\ref{tab:yields}, for all three \shat~measures.

\begin{table}[h]
\begin{centering}
\begin{tabular}{l | ll}
\hline
& Signal region [GeV] & Sideband region [GeV] \\
\hline
\mx, \ma = 350 GeV, 40 GeV & [325, 375) & [275, 325) $\cup$ [375, 425) \\
\mx, \ma = 700 GeV, 100 GeV & [675, 725) & [625, 675) $\cup$ [725, 775) \\
\hline
\end{tabular}
\caption{Table of selections on \shat~that define the analysis signal and sideband regions.\label{tab:regions}}
\end{centering}
\end{table}

\begin{table}[h]
\begin{centering}
\begin{tabular}{l|ll|ll}
\hline
& & & Signal Region & Sideband Region \\
\hline
\multirow{2}{*}{Truth \shat}
& \multirow{2}{*}{m$_X$ = 350 GeV} & Background & 12133 & 25000 \\ & & Signal & 76 & 0 \\ 
& \multirow{2}{*}{m$_X$ = 700 GeV} & Background & 10832 & 22407 \\ & & Signal & 68 & 0 \\ 
\hline
\multirow{2}{*}{$\gamma$-measured \shat}
& \multirow{2}{*}{m$_X$ = 350 GeV} & Background & 12121 & 25000 \\ & & Signal & 76 & 0 \\ 
& \multirow{2}{*}{m$_X$ = 700 GeV} & Background & 10441 & 21928 \\ & & Signal & 65 & 0 \\ 
\hline
\multirow{2}{*}{Hadron-measured \shat}
& \multirow{2}{*}{m$_X$ = 350 GeV} & Background & 12147 & 25000 \\ & & Signal & 76 & 2 \\ 
& \multirow{2}{*}{m$_X$ = 700 GeV} & Background & 11754 & 24459 \\ & & Signal & 74 & 1 \\ 
\hline
\end{tabular}
\caption{Training yields for all \shat~measures, using the region definitions given in Table~\ref{tab:regions}, for an example signal injection corresponding to a 1.0$\sigma$ significance.\label{tab:yields}}
\end{centering}
\end{table}

Two different training configurations for classification are used, which vary in how the two input labeled classes are populated. 

\begin{itemize}
    \item A semi-supervised scenario is studied as a benchmark. In this case, a neural network is trained to classify background events from a combination of signal and background events in the signal region, where the amount of signal contamination varies. In a context where real \ee~collision data is available, this would correspond to classifying background-only simulation from data (where a potential signal could be included)~\cite{DAgnolo:2018cun,DAgnolo:2019vbw}. The assumption is made that the simulation perfectly describes the data; though this is not achievable with current simulations, future advancements may make such a strategy feasible.

    \item Weakly supervised learning can be used to enhance the presence of potential localized signals in data, without relying on simulation. Here, the classification without labels (CWoLa) approach is used, where a classifier is trained to distinguish events in the signal region from those in the sidebands. When a signal is present, this method takes advantage of the different compositions in signal and sideband regions to learn how to distinguish background from signal~\cite{Metodiev:2017vrx,Collins:2018epr,collaboration2020dijet,Collins:2019jip}.
\end{itemize}
The envisioned analysis strategy in real data would be to define a signal-enriched region using the classifier score, and then estimate the background with a completely data-driven bump hunt (e.g. a parametric fit to the sidebands). 

In both the semi-supervised and weakly-supervised scenarios, classifiers are parameterized with a Deep Sets model~\cite{10.5555/3294996.3295098}, using the Particle Flow Networks (PFN) implementation~\cite{Komiske:2018cqr} in the \texttt{EnergyFlow} package\footnote{\url{https://energyflow.network/}.}. A PFN is composed of a per-jet\footnote{We also investigated using all particle-flow candidates directly instead of jets, but the sensitivity was worse than the jet-based learning.  Further details are provided in Sec.~\ref{sec:results}.} function $\Phi$ and an event-level function $F$, both learnable and each with 3 layers and 20 nodes per layer. The inputs to the PFN are the variable-length set of jets, considering up to 15 jets per event. For each jet, 10 variables are considered: its four-vector, 5 angular radiation moments ($N$-subjettiness~\cite{Thaler:2010tr,Thaler:2011gf} for $1\leq N\leq 5$) and a flavor-tagging discriminant encoding 4 bins of $b$-tagging efficiency (50\%, 70\%, 90\%, 100\%). Plots of the input variables for leading and sub-leading jets in signal and background processes are shown in Appendix~\ref{app:pfnvars}.

The neural networks are trained using \textsc{keras}~\cite{chollet2015keras} with the \textsc{tensorflow}~\cite{tensorflow2015-whitepaper} backend. The categorical cross-entropy loss function was minimized, using the Adam optimizer~\cite{adam} with an initial learning rate of 0.0001. Starting with higher (up to 0.01) and lower (down to 10$^{-5}$) learning rates was found to be suboptimal. Adagrad~\cite{adagrad} and RMSProp~\cite{rmsprop} optimizers were also studied, with no significant impact on the performance. The PFN was trained for 30 epochs with a batch size of 100. A longer training time of 100 epochs was also considered and did not strongly affect the final performance. 

\begin{figure}
\begin{centering}
\includegraphics[scale=0.4]{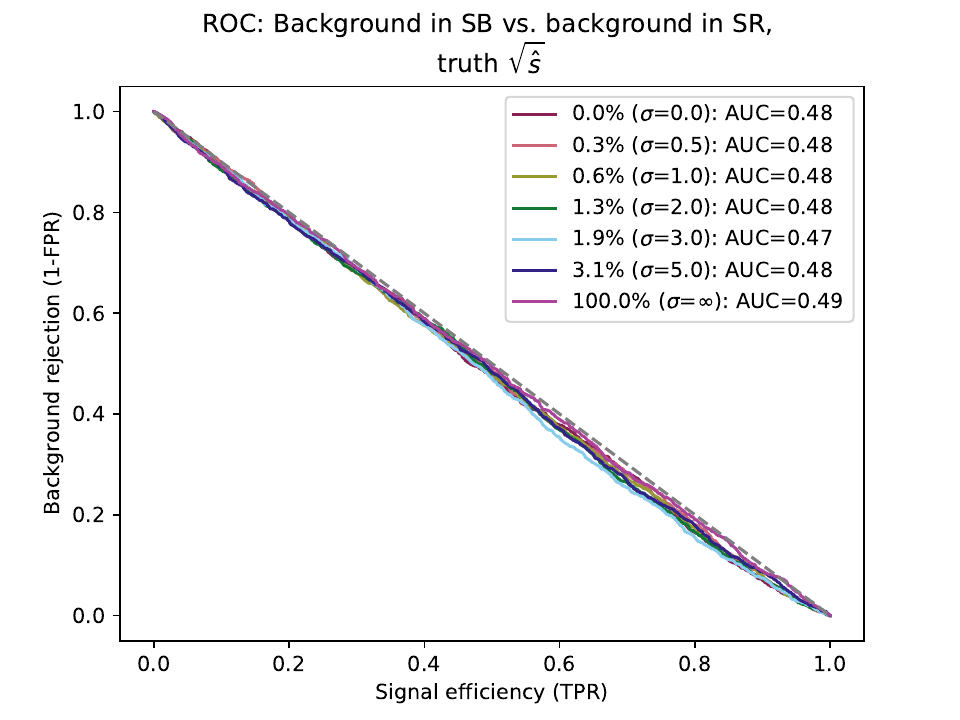}
\includegraphics[scale=0.4]{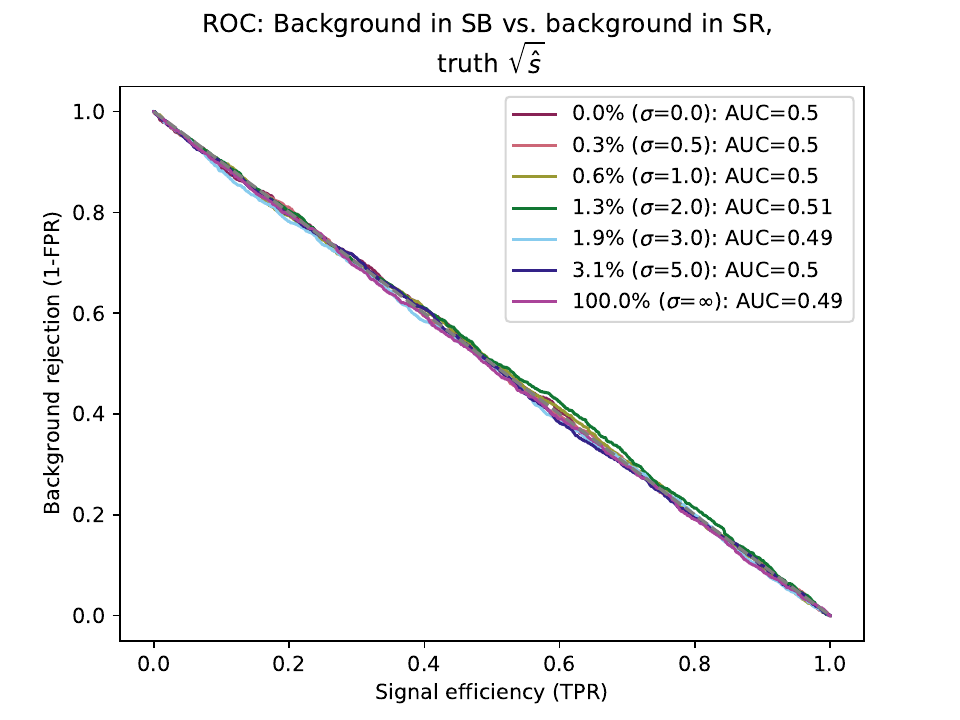}
\caption{Receiver operating characteristic (ROC) curves from the weakly supervised training setup, comparing background in the sideband vs. background in the signal region, for the $m_{X}=350$~GeV signal region on the left and the $m_{X}=700$~GeV signal region on the right. Each curve represents a different signal contamination in the training set.\label{fig:pfn_truth_BvsB}}
\end{centering}
\end{figure}

In the weakly supervised scenario where training utilizes events from different bins of \shat, care must be taken to ensure that the network output is agnostic to the \shat~of the events. A per-event normalization procedure is implemented to mitigate the \shat~correlation. Each jet's $\eta$ and $\phi$ is centered on the average value for all jets in the event, and its $p_T$ is scaled by the sum of jet transverse momentum in the event. The efficacy of the normalization procedure is verified by training the network to identify background events in the signal region from background events in the sideband. Since these events should only vary in their \shat~values, the normalization procedure can be deemed functional if the classifier is unable to discern these two classes of background events. Figure~\ref{fig:pfn_truth_BvsB} shows the result of this training, confirming that the chosen normalization is sufficiently able to remove significant correlations of learned information with \shat.

Considerable variance in performance was observed across models with identical training scenarios. An ensemble procedure was developed to mitigate the effect of these fluctuations. Each training result presented here represents the average of 50 trained models, each with a random signal injection. The predicted values on the test set are subject to a non-linear transform known as quantile scaling, after which they follow a normal distribution. Models are then combined by averaging the quantile-scaled results for all 50 models.  The results should therefore be interpreted as the expected/average sensitivity.


%% file: sections/results.tex
\section{Results}
\label{sec:results}

The results of the training are displayed in two forms. The first is the receiver operating characteristic curve (ROC), which shows background rejection (inverse of the false positive rate, FPR) as a function of the signal efficiency (true positive rate, TPR), and demonstrates the discriminating power of the output net score. Additionally, the significance improvement characteristic (SIC, TPR/$\sqrt{\text{FPR}}$) is provided, which shows the signal sensitivity as a function of signal efficiency.  

The SIC can be used as a proxy of how the output network score can enhance a signal excess in a physics analysis context.  In particular, if $1\sigma$ of signal is injected and the SIC value at a particular signal efficiency is $n$, then a cut on the classifier corresponding to that efficiency is expected to increase the amount of signal to about $n\sigma$ (local significance). We do not perform a background fit, so the signal and background yield used to compute the SIC are determined by counting the number of events of each type in the signal region after a particular cut on the classifier output. In practice, the actual significance would slightly degrade from uncertainties in the fit, but as with other recent papers on anomaly detection in HEP, we decouple the network performance and the background fit quality in order to focus on the classifier methodology.

\input{sections/pfn}

\input{sections/eventlevel}

%% file: sections/pfn.tex
\subsection{Particle Flow Networks}
\label{subsec:pfn}

\begin{figure}
\begin{centering}
\includegraphics[scale=0.4]{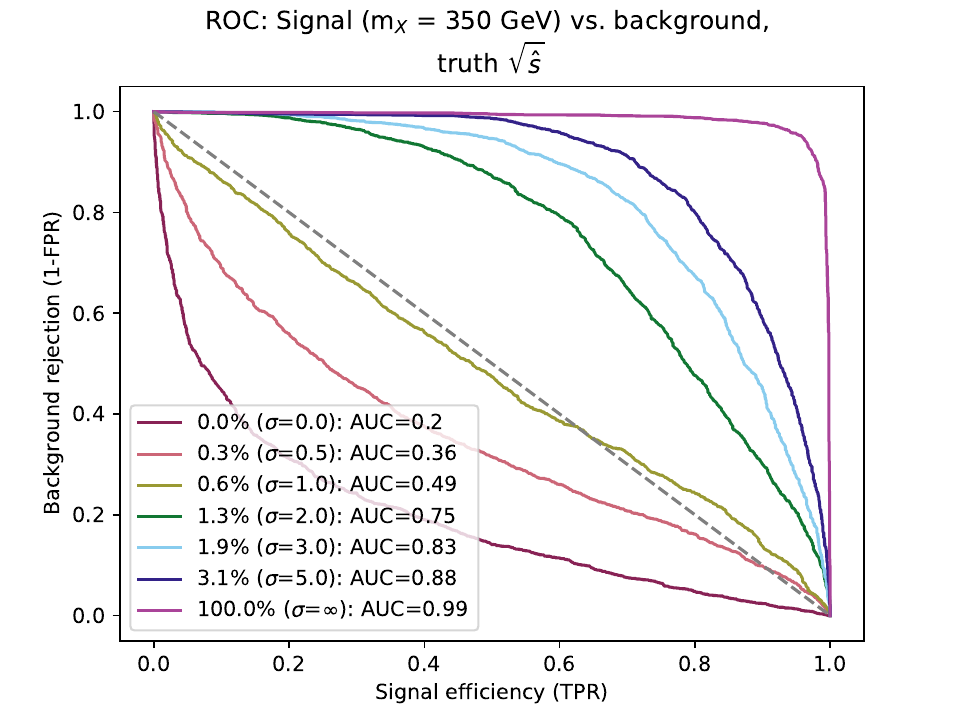}
\includegraphics[scale=0.4]{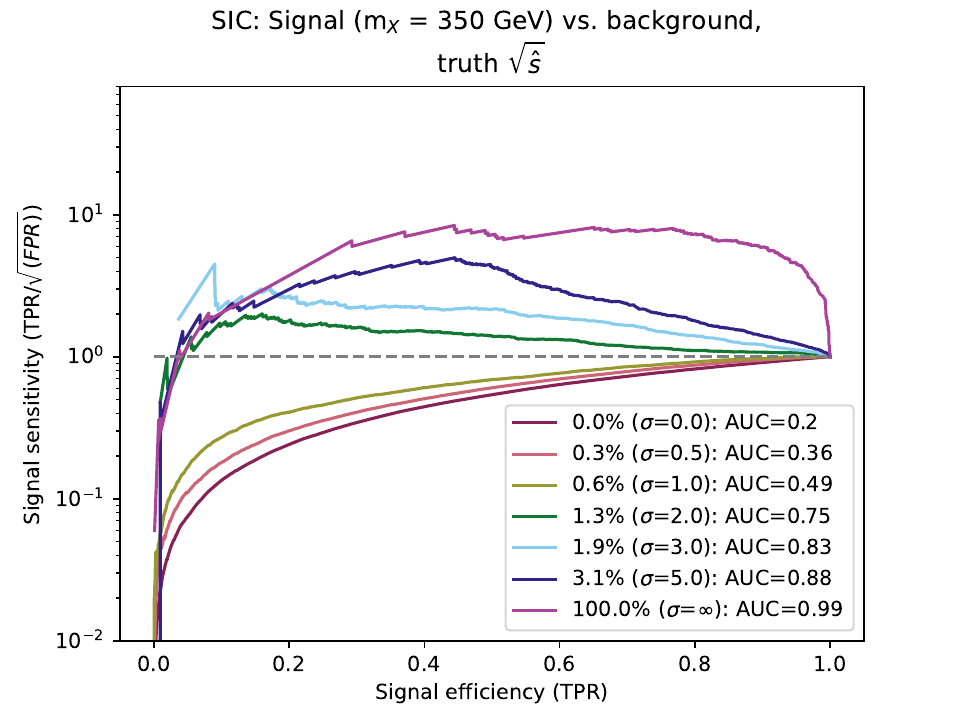}
\includegraphics[scale=0.4]{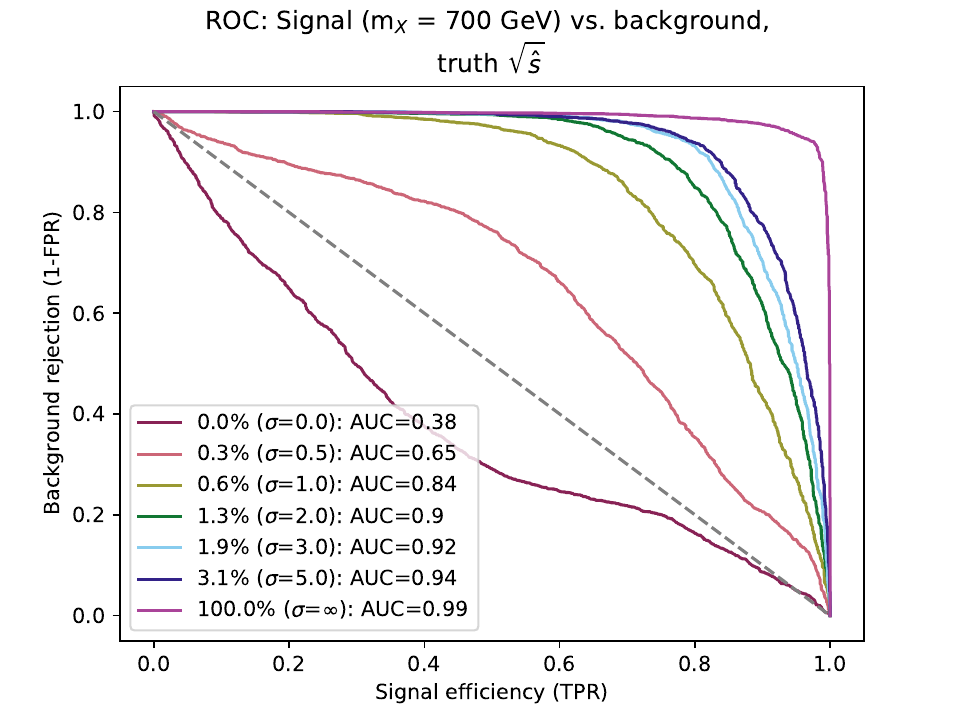}
\includegraphics[scale=0.4]{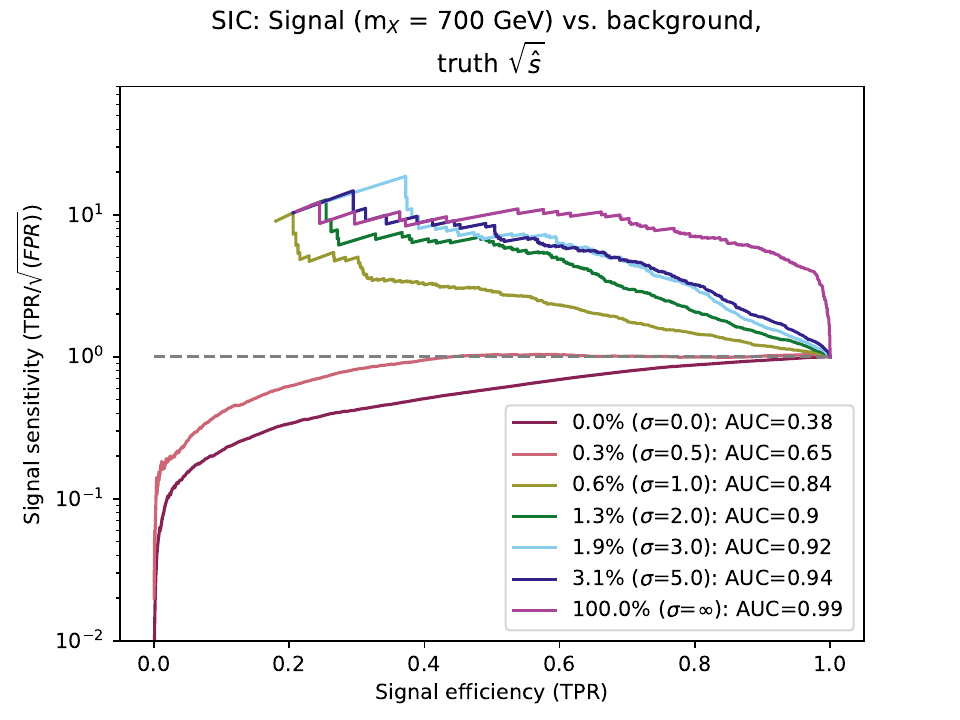}
\caption{Semi-supervised training results in the form of ROC (left) and SIC (right) curves for two signals, m$_{X}$ = 350 GeV (top) and m$_{X}$ = 700 GeV (bottom)  vs. background. The dashed grey line indicates no improvement from the use of the neural network score.  A significance of $\infty$ indicates the case of pure signal versus pure background. \label{fig:pfn_benchmark_roc}}
\end{centering}
\end{figure}

The semi-supervised training results are discussed first, as they provide a benchmark of the ideal classifier performance when separating signal from background in the signal region only. Figure~\ref{fig:pfn_benchmark_roc} gives the ROC and SIC curves for both signal mass hypotheses as a result of the semi-supervised training configuration. The 0\% signal contamination curve is shown as a reference. In this figure and in subsequent results, the lack of signal leads to a network that learns arbitrary discriminating information about the training inputs, resulting in small deviations of the area-under-curve (AUC) value from that of a completely random classifier (which can lead to an apparently `anti-tagging'). A similar trend in the 0.3\% contamination result indicates that this amount of signal is consistently too small for the network to reliably learn and distinguish it from background. 

With semi-supervised training, the network is able to detect an $m_X=700$~GeV signal contamination of 0.6\% from background in a single bin of \shat~and increase its significance by a factor of 11. The setup is considerably less sensitive to the $m_X=350$~GeV signal, where the 0.6\% contamination is indistinguishable, but doubled significance is achieved for a 1.3\% contamination. Details on the impact of the neural net output selection can be found in Table~\ref{tab:resultsummary}. Both of these signal injections correspond to excesses of 2$\sigma$ sensitivity or less, making the role of the neural network substantial in elevating potential new physics to the discovery level of sensitivity. 

\begin{figure}
\begin{centering}
\includegraphics[scale=0.4]{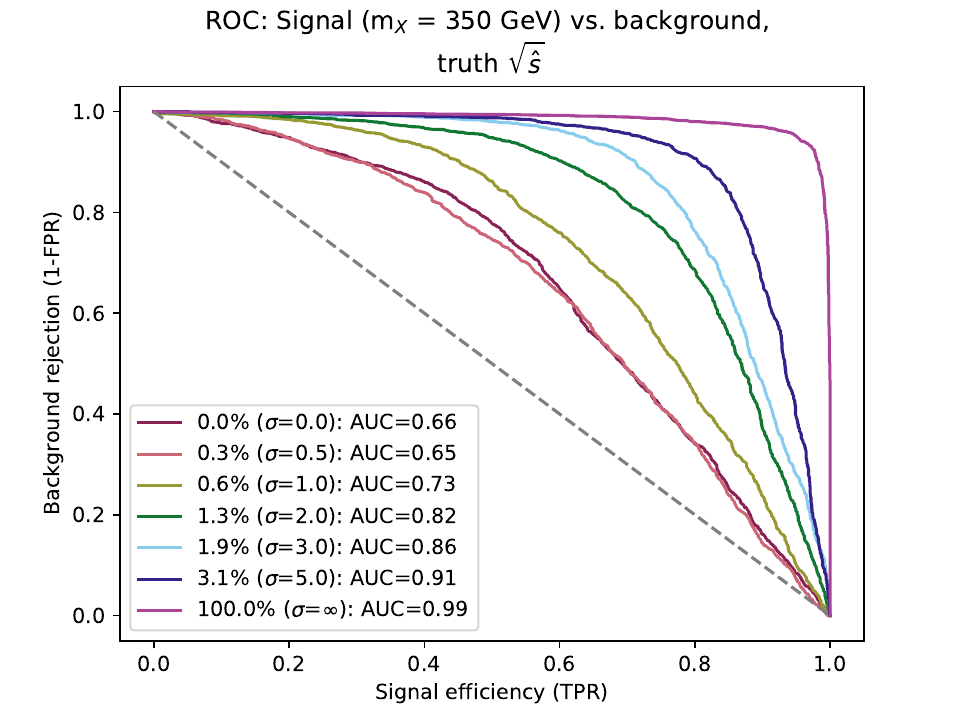}
\includegraphics[scale=0.4]{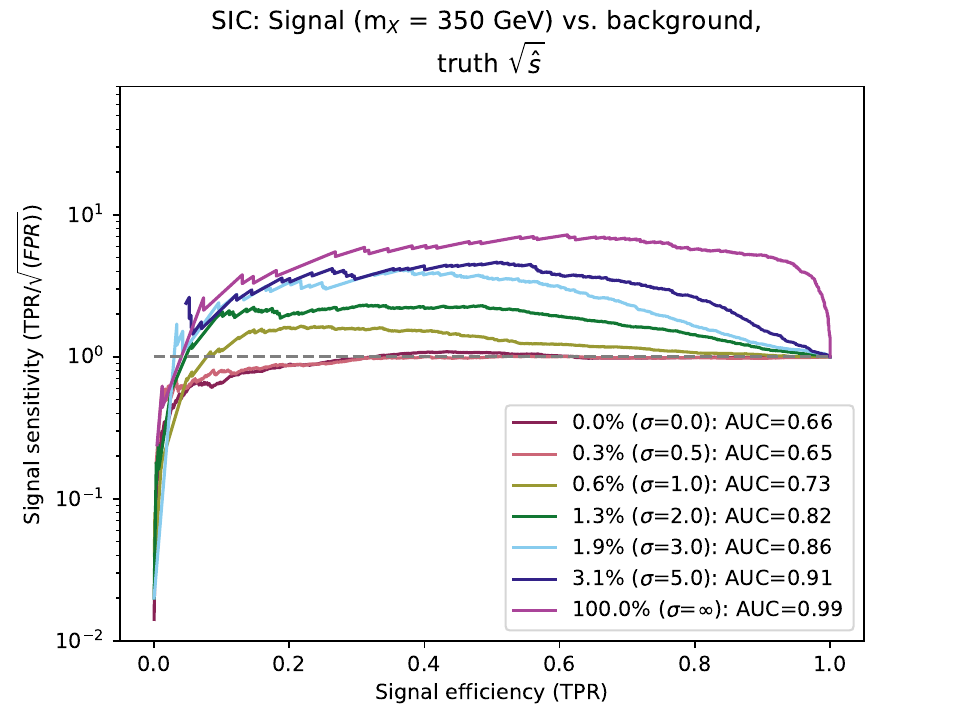}
\includegraphics[scale=0.4]{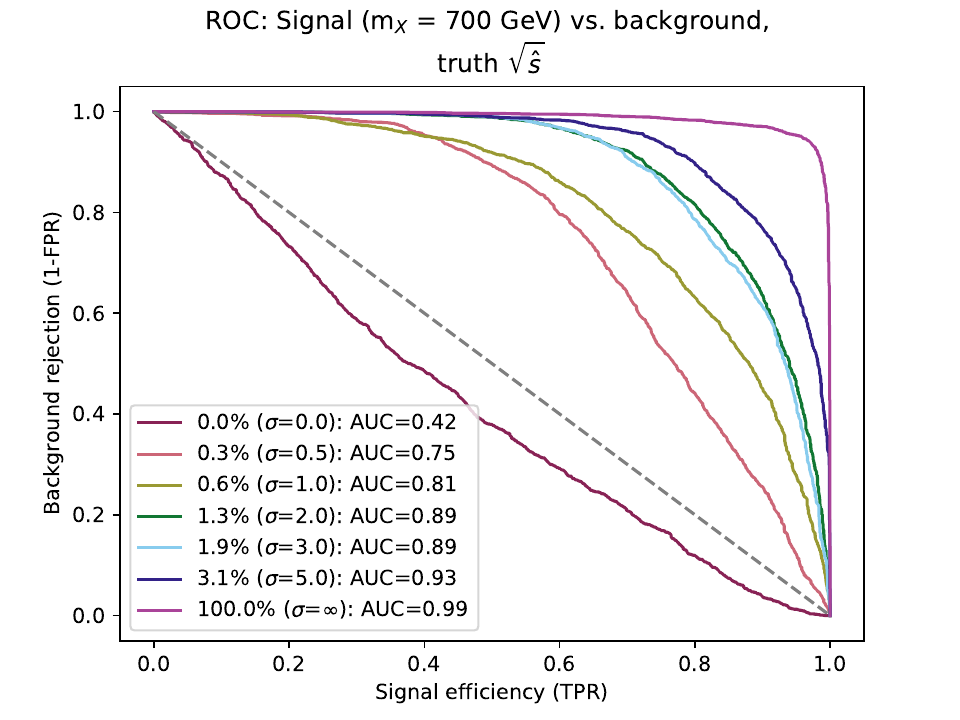}
\includegraphics[scale=0.4]{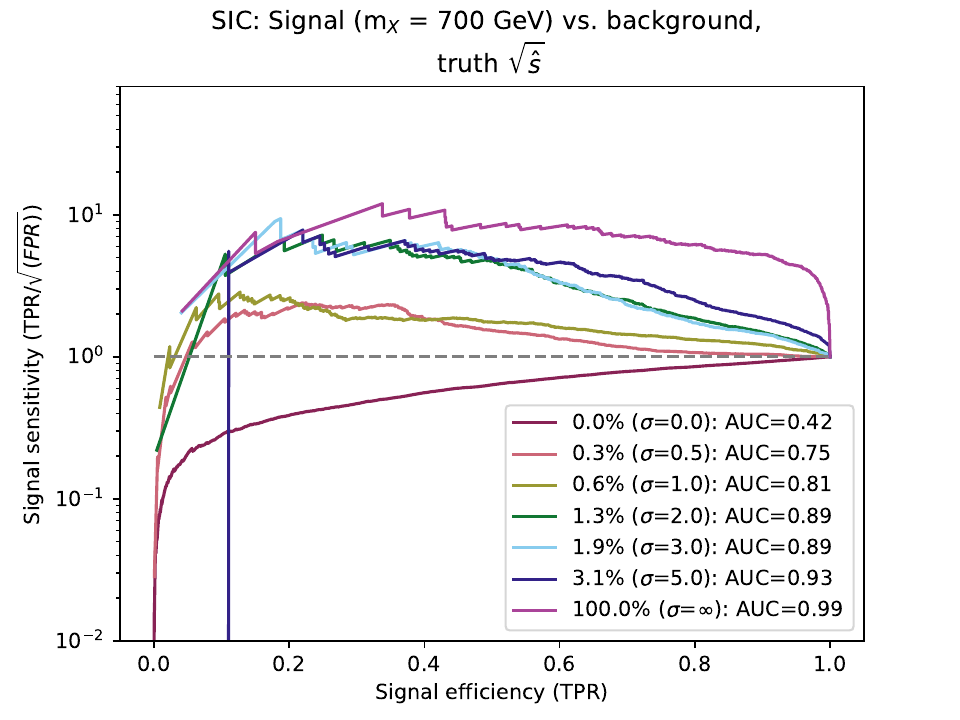}
\caption{Weakly supervised training results in the form of ROC (left) and SIC (right) curves for two signals, $m_{X} = 350$~GeV (top) and $m_{X} = 700$~GeV (bottom)  vs. background. The dashed grey line indicates no improvement from the use of the neural network score.  A significance of $\infty$ indicates the case of pure signal versus pure background.\label{fig:pfn_truth_roc}}
\end{centering}
\end{figure}

Figure~\ref{fig:pfn_truth_roc} shows the training results for the weakly supervised training scenario for both signal mass hypotheses.  The network is similarly able to enhance a signal contamination down to 0.6\% for both signal mass hypotheses, again with specific sensitivity increase factors given in Table~\ref{tab:resultsummary}. The comparable results in Figures~\ref{fig:pfn_benchmark_roc} and~\ref{fig:pfn_truth_roc} show that differences in the background-only case between the signal region and sideband region are not reducing the signal sensitivity. 


An even higher level of input complexity is possible through utilizing the PFN architecture with all particles in the event. This representation was tried in the training, but it was found to be less performant than the jet-level PFN studied here. This indicates a clear dependency of classifier performance on both input dimensionality and accordingly the amount of input statistics available in training.

%% file: sections/eventlevel.tex
\subsection{Event Level}
\label{subsec:eventlevel}

To contextualize the results from the PFN training, we perform the same weakly supervised training procedure over input events described solely by event-level kinematic quantities and event shape variables. 

The event-level network setup also utilized the \texttt{EnergyFlow} package, but replaced the PFN architecture with a dense neural network (DNN) with 15 input dimensions and two dense layers with 100 nodes each. A dropout rate of 0.2 was added to mitigate observed overtraining. As with the PFN, the DNN was trained for 30 epochs with a batch size of 100. Fifteen variables were used to describe input events: the leading and subleading large-radius jet masses and transverse momenta, the leading photon \pt, the X particle \pt, the particle and jet multiplicity, the ratio of leading jet and X \pt~to leading photon pt, ln$y_{23}$, aplanarity, sphericity, transverse sphericity, and total jet mass. Distributions of these variables for signal and background events can be found in Appendix~\ref{app:eventlevelvars}. 

\begin{figure}
\begin{centering}
\includegraphics[scale=0.4]{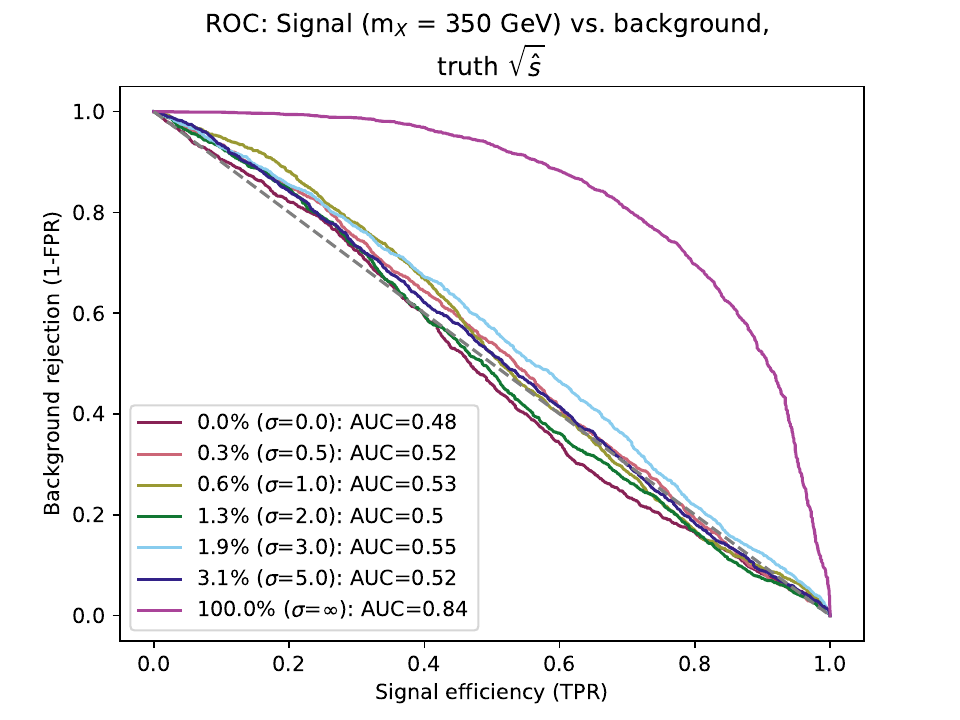}
\includegraphics[scale=0.4]{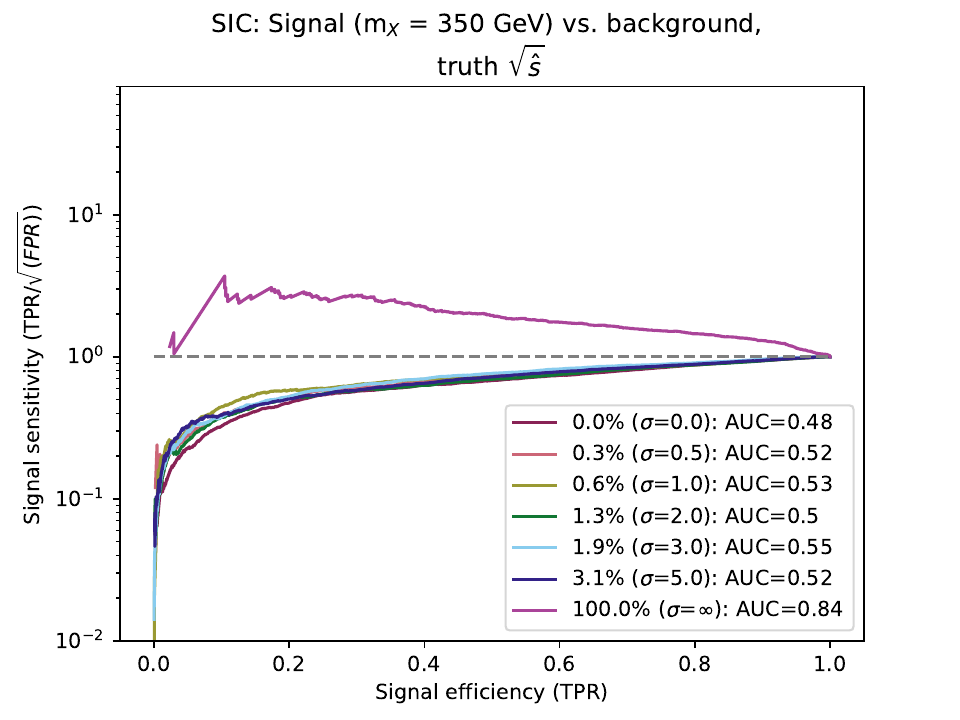}
\includegraphics[scale=0.4]{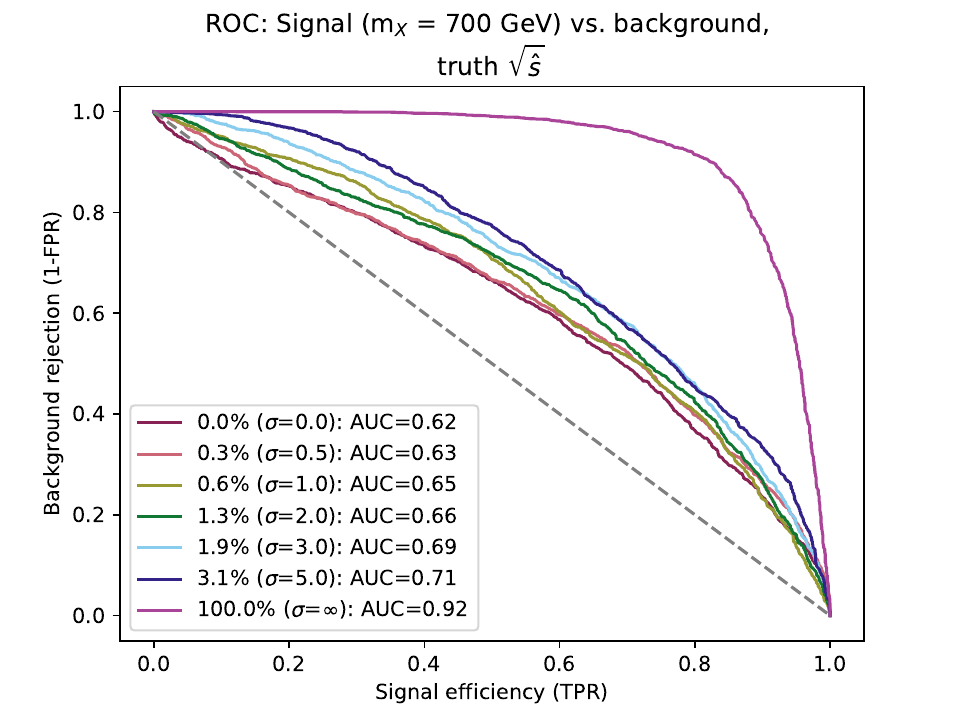}
\includegraphics[scale=0.4]{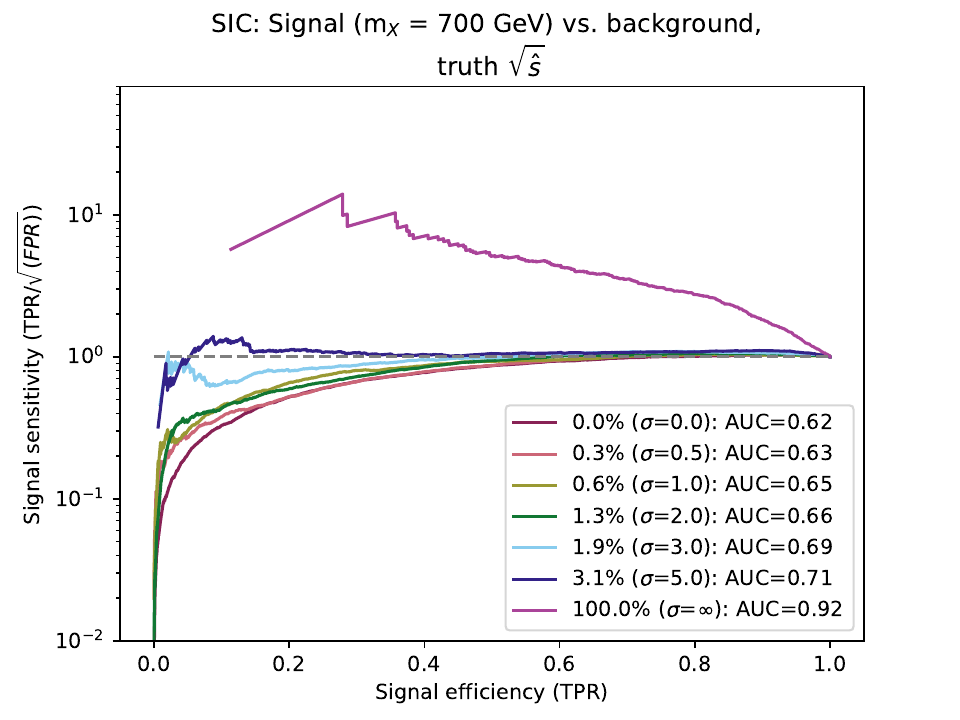}
\caption{Weakly supervised training results using event-level input variables, in the form of ROC (left) and SIC (right) curves for $m_{X} = 350$~GeV (top) and $m_{X} = 700$~GeV (bottom). The dashed grey line indicates no improvement from the use of the neural network score.   A significance of $\infty$ indicates the case of pure signal versus pure background.\label{fig:evt_truth_roc}}
\end{centering}
\end{figure}

Results are shown in Figure~\ref{fig:evt_truth_roc} for the weakly supervised training scenario. The inability of the DNN to distinguish signal from background, except in the 100\% signal contamination scenario, indicates that the event-level variables are suboptimal for the signal of this study. Comparison to Figure~\ref{fig:pfn_truth_roc}, which gives the analogous result for the PFN training, demonstrates the benefit of using high-dimensional input representations for the task of anomaly detection. Comparable signal sensitivity is delivered by a fully supervised signal vs. background training on event-level variables, and a PFN weakly supervised training with only 3.1\% signal contamination. 

%% file: sections/futuredetector.tex
\section{Future Detector Considerations}
\label{sec:futuredetector}

To extrapolate these results to a search in real collision data, the same method is applied using regions defined with a measured \shat~instead of one computed with truth-level quantities. 
Two different methods for measuring the total available energy are considered.
One assumes that the ISR photon is captured by the detector, and therefore uses the measurement of its energy subtracted from the incoming electron-positron \shat~as a proxy for the amount of energy available in the collision.
This is referred to as the \textit{photon-measured \shat}.
The second is the \textit{hadron-measured \shat}, which covers the scenario where the photon is lost and the collision CoM must be obtained through measurements of the final-state hadrons.  Note that the highest $p_T$ photon is always used for these calculations.  In the photon-measured case, if the true ISR photon is out of acceptance, the predicted \shat~will be significantly different from the true one.  In the hadron-measured case, the selected photon is excluded from the calculation of \shat.

\begin{figure}
\begin{centering}
\includegraphics[width=0.45\textwidth]{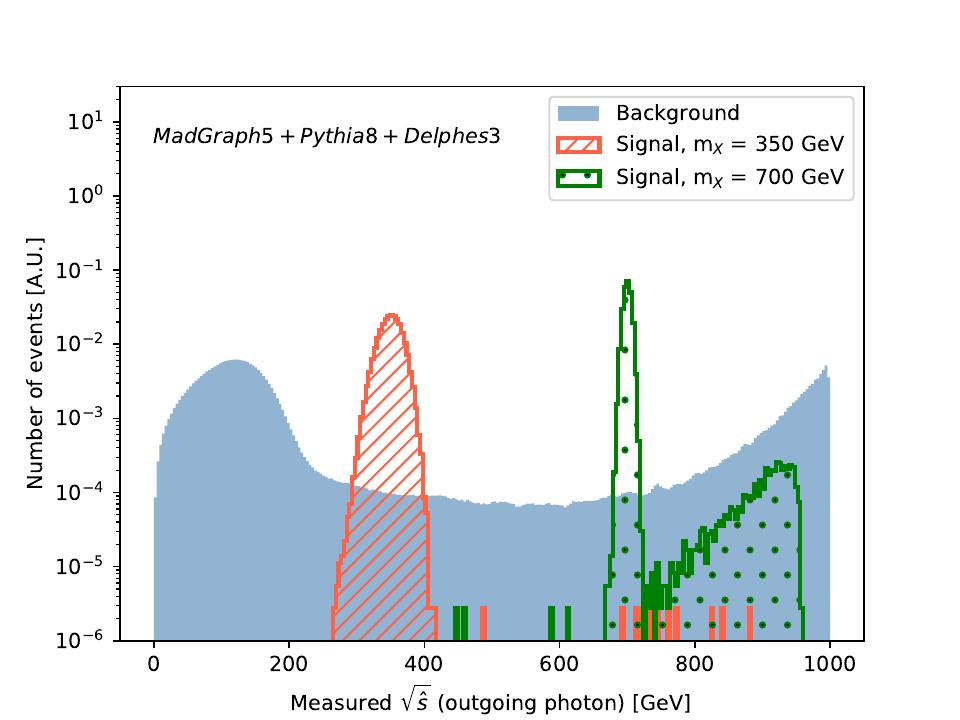}
\includegraphics[width=0.45\textwidth]{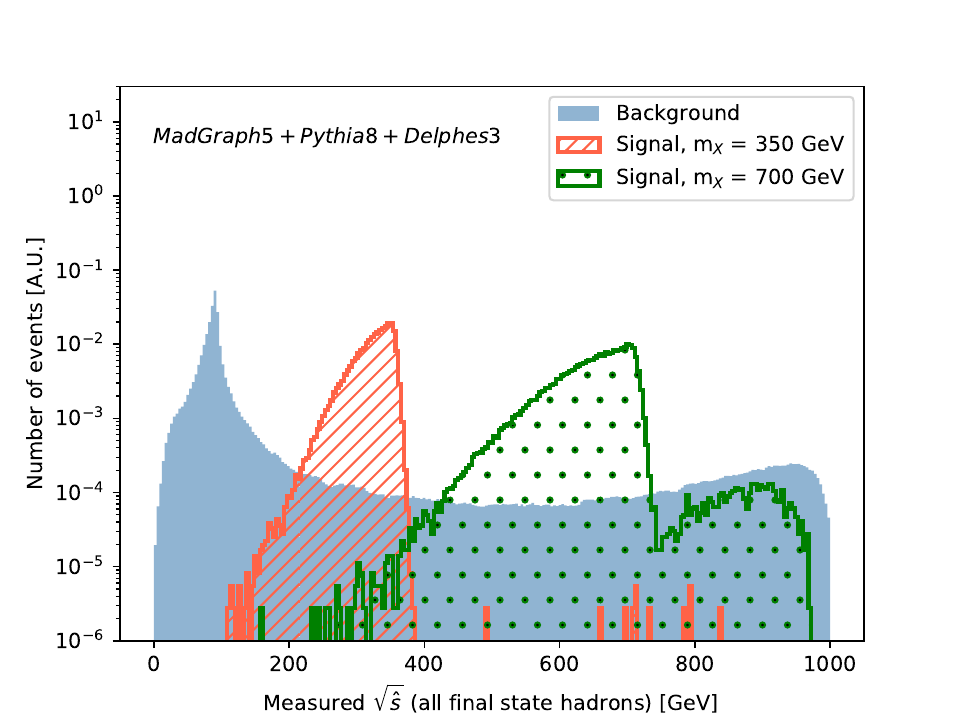}
\caption{Distributions of the measured collision $\sqrt{\hat{s}}$ when the outgoing photon is captured and subtracted from the initial collision energy (left), and computed using only the final state hadrons in the event (right) \label{fig:measuredsqrtshat}}
\end{centering}
\end{figure}

Figure~\ref{fig:measuredsqrtshat} shows distributions of these two $\sqrt{\hat{s}}$ measurements for the background and both signal hypotheses. The incorporation of detector information gives each resonance a non-negligible width due to smearing introduced by detector resolution. As a result, the signal-to-noise in the signal region is lower. As seen in Table~\ref{tab:yields}, this width can also create some signal contamination in the sideband. Both of these effects make the discrimination task more challenging.  In the photon-measured case, the signal and $Z$ peaks are approximately symmetric, with the width dominated by the photon energy resolution.   The high-\shat~tail in the 750 GeV case is the result of events where the true ISR photon is out of acceptance and a random photon (the next highest $p_T$ one) is used to compute \shat.  In the hadron-measured case, the signal peaks are asymmetric because there are both resolution and acceptance effects playing a role.  The $Z$ peak is sharper for the hadron-measured case compared with the photon-measured case because the absolute energy resolution is better at low $p_T$: for the hadron-measured case, all of the particles are $\lesssim m_Z$ while for the photon-measured case, the photon energy is nearly $\sqrt{s}$.


\begin{figure}
\begin{centering}
\includegraphics[scale=0.4]{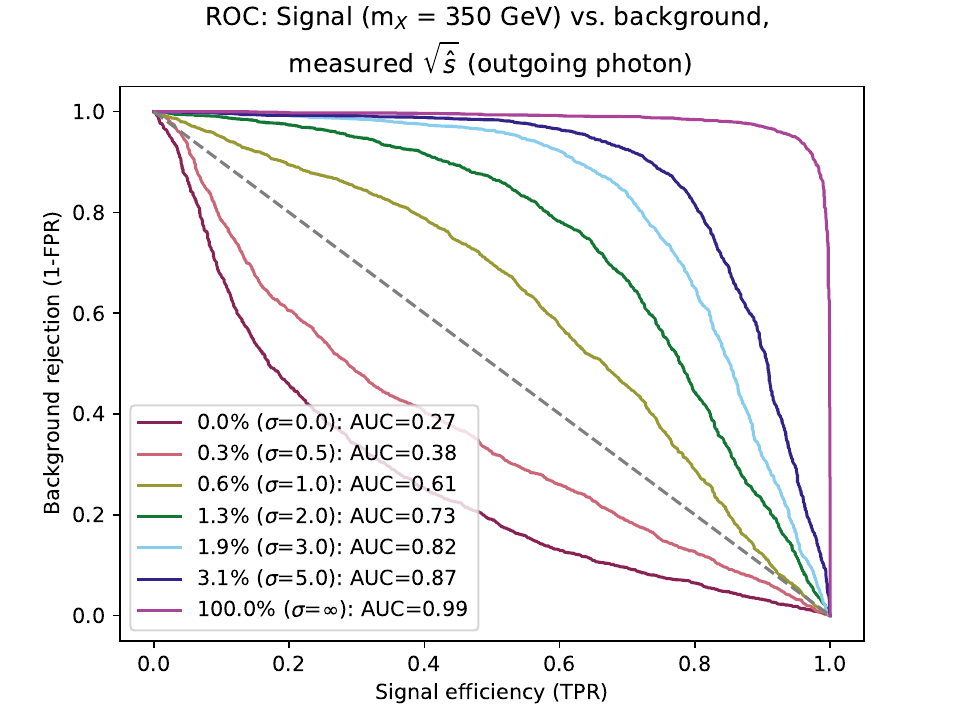}
\includegraphics[scale=0.4]{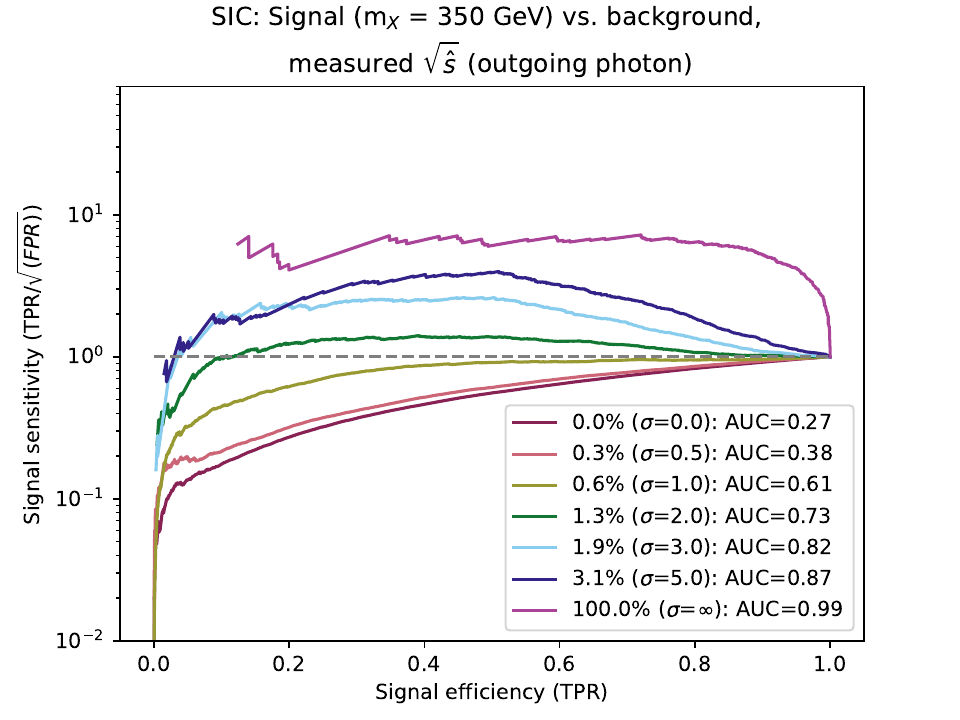}
\includegraphics[scale=0.4]{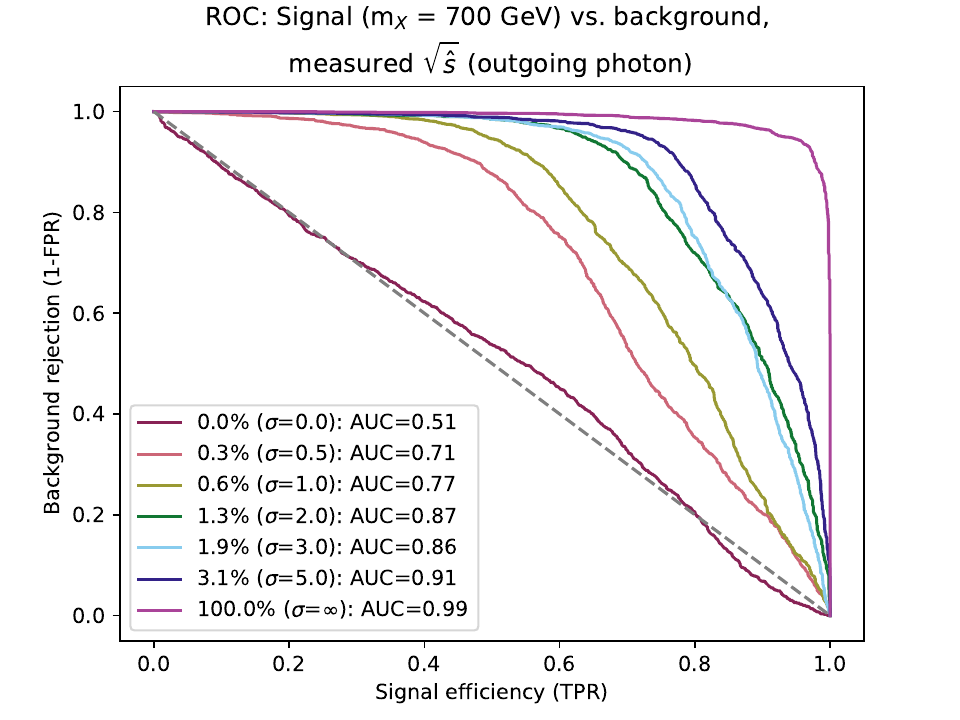}
\includegraphics[scale=0.4]{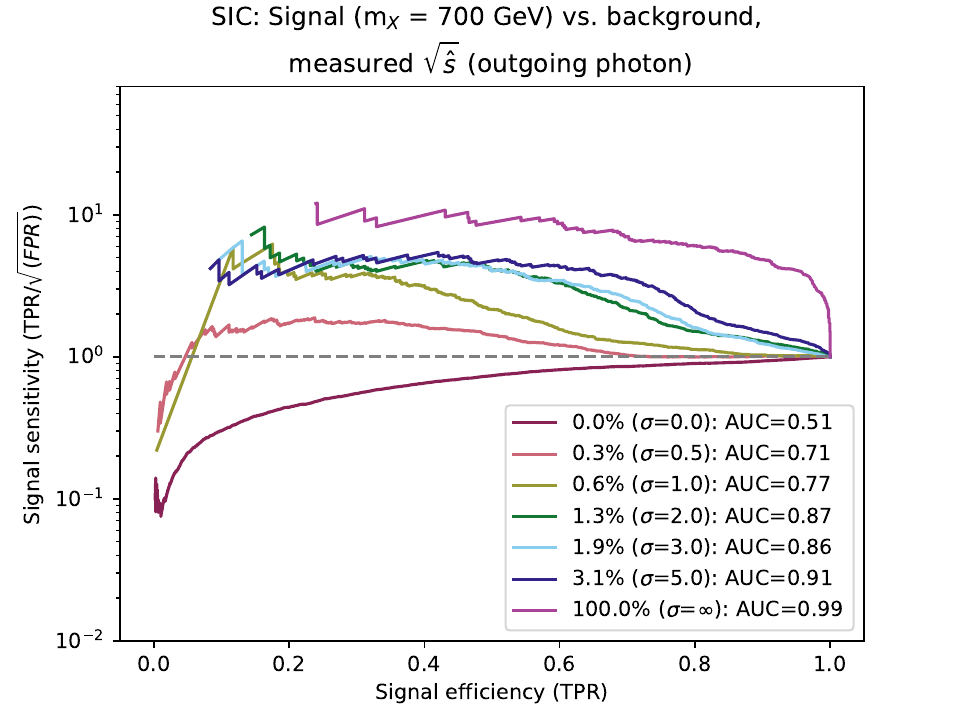}
\caption{Weakly supervised PFN training results using photon-measured \shat, in the form of ROC (left) and SIC (right) curves for $m_{X} = 350$~GeV (top) and $m_{X} = 700$~GeV (bottom) vs. background. The dashed grey line indicates no improvement from the use of the neural network score.   A significance of $\infty$ indicates the case of pure signal versus pure background.\label{fig:pfn_measph_roc}}
\end{centering}
\end{figure}

\begin{figure}
\begin{centering}
\includegraphics[scale=0.4]{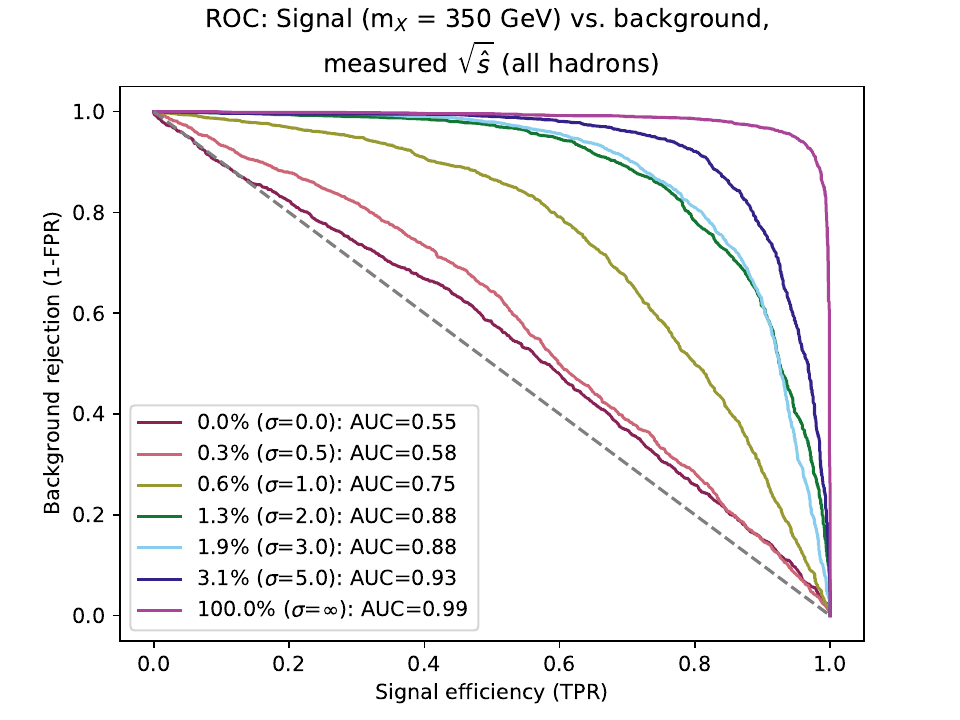}
\includegraphics[scale=0.4]{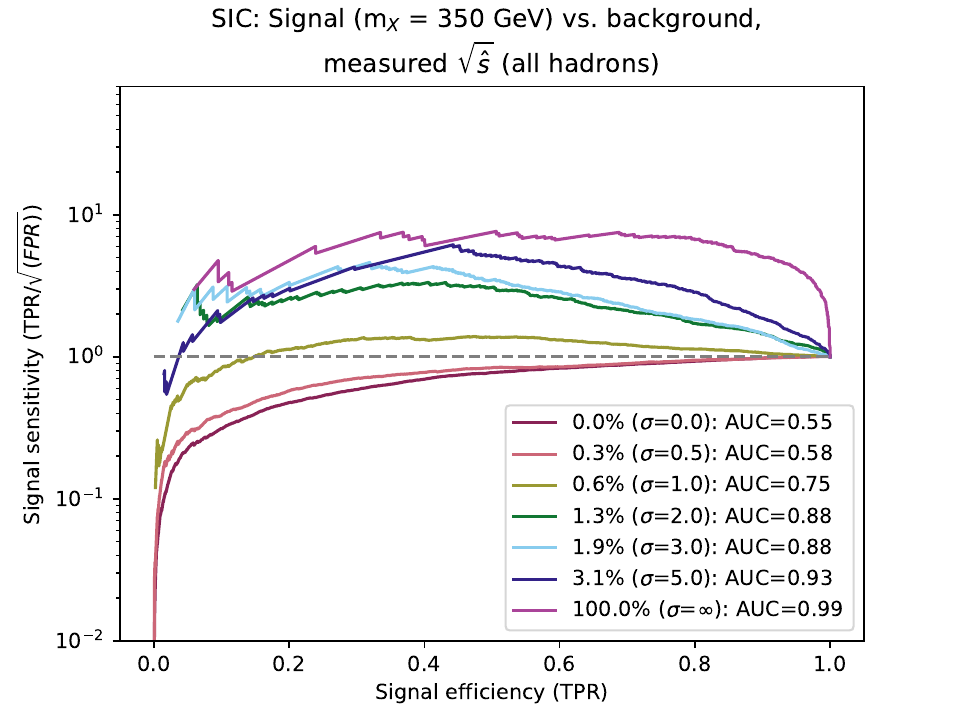}
\includegraphics[scale=0.4]{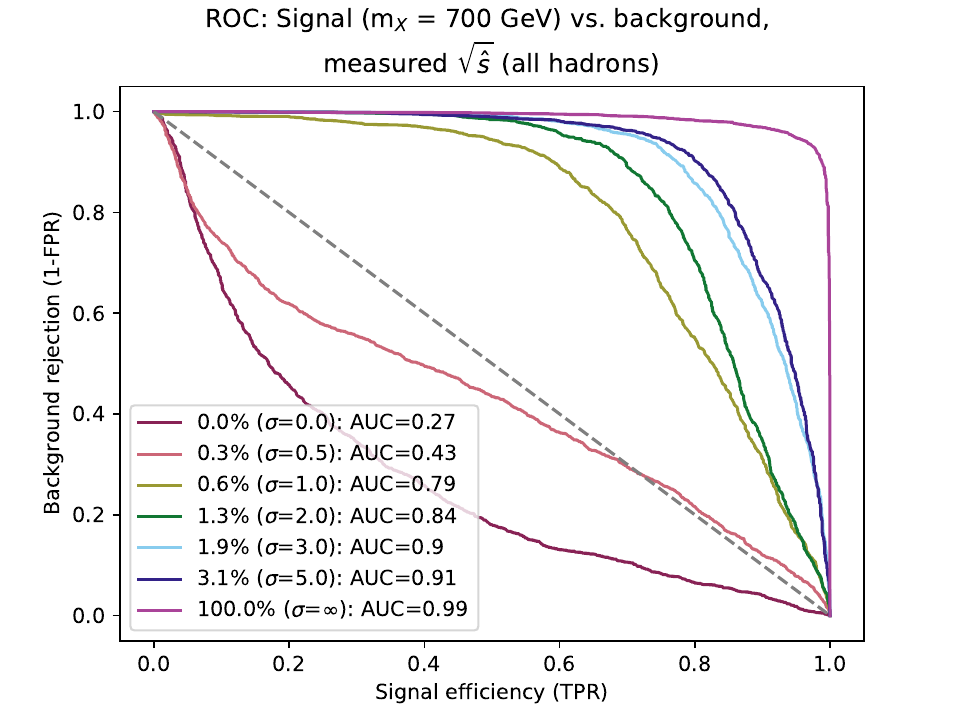}
\includegraphics[scale=0.4]{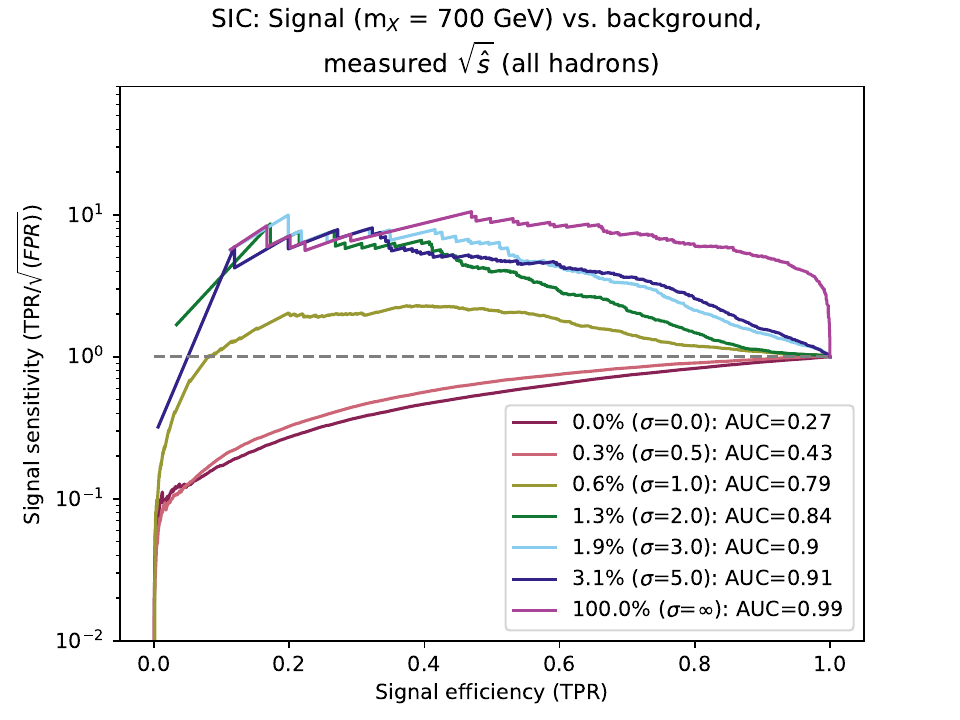}
\caption{Weakly supervised PFN training results using hadron-measured \shat, in the form of ROC (left) and SIC (right) curves for $m_{X} = 350$~GeV (top) and $m_{X} = 700$~GeV (bottom) vs. background. The dashed grey line indicates no improvement from the use of the neural network score.   A significance of $\infty$ indicates the case of pure signal versus pure background.\label{fig:pfn_meas_roc}}
\end{centering}
\end{figure}

Performance of the method can be found in Figures~\ref{fig:pfn_measph_roc} and~\ref{fig:pfn_meas_roc}, for the photon-measured and hadron-measured \shat, respectively. The signal significance is calculated using the signal region that is defined by the \shat~measure of interest, and is not normalized to the original truth sensitivity. Although the sensitivity is generally diminished by detector effects, there is still strong enhancement for a variety of signal injections, representing potential for this method in real collision data. 
Future innovations on hardware (e.g. increased acceptance) and software (e.g. combining photon- and hadron-measurements) may be able to close any remaining gaps between the truth \shat~and the reconstructed version(s).

These studies also indicate the effectiveness of the method in a real analysis context where a sliding window would be used to define sets of signal regions and sidebands across \shat. In this case, there would necessarily be some sidebands with more signal events than the signal region, leading to suboptimal sensitivity, though at least one such region will exist where the signal is mostly in the signal region.

A summary of all training configurations and the impact of the neural net on the achievable signal sensitivity can be found in Table~\ref{tab:resultsummary}. 

\begin{table}[h]
\begin{centering}
\resizebox{\textwidth}{!}{
\begin{tabular}{p{2cm}|lllllllll}
\hline
& & Signal Signif. [$\sigma$] & 0 & 0.5 & 1.0 & 2.0 & 3.0 & 5.0 & $\infty$ \\
\hline
\multirow{2}{2cm}{Semi-supervised, truth \shat}
& \multirow{2}{*}{m$_X$ = 350 GeV} & AUC & 0.2 & 0.36 & 0.49 & 0.75 & 0.83 & 0.88 & 0.99 \\ & & Max SIC (new $\sigma$) & 1.0 (0.0) & 1.0 (0.5) & 1.0 (1.0) & 2.0 (4.0) & 4.5 (14) & 5.0 (25) & 8.4 ($\infty$) \\ 
& \multirow{2}{*}{m$_X$ = 700 GeV} & AUC & 0.38 & 0.65 & 0.84 & 0.9 & 0.92 & 0.94 & 0.99 \\ & & Max SIC (new $\sigma$) & 1.0 (0.0) & 1.0 (0.5) & 10.2 (10.2) & 12.7 (25) & 18.6 (56) & 14.7 (74) & 12.3 ($\infty$) \\ 
\hline

\multirow{2}{2cm}{Weakly supervised, truth \shat}
& \multirow{2}{*}{m$_X$ = 350 GeV} & AUC & 0.66 & 0.65 & 0.73 & 0.82 & 0.86 & 0.91 & 0.99 \\ & & Max SIC (new $\sigma$) & 1.0 (0.0) & 1.0 (0.5) & 1.6 (1.6) & 2.3 (4.6) & 4.1 (12) & 4.6 (23) & 7.2 ($\infty$)  \\ 
& \multirow{2}{*}{m$_X$ = 700 GeV} & AUC & 0.42 & 0.75 & 0.81 & 0.89 & 0.89 & 0.93 & 0.99 \\ & & Max SIC (new $\sigma$) & 1.0 (0.0) & 2.4 (1.2) & 2.9 (2.9) & 7.2 (14) & 9.4 (28) & 7.8 (39) & 12.0 ($\infty$) \\ 
\hline

\multirow{2}{2cm}{Weakly supervised, photon \shat}
& \multirow{2}{*}{m$_X$ = 350 GeV} & AUC & 0.27 & 0.38 & 0.61 & 0.73 & 0.82 & 0.87 & 0.99 \\ & & Max SIC (new $\sigma$) & 1.0 (0.0) & 1.0 (0.5) & 1.0 (1.0) & 1.4 (2.8) & 2.6 (7.8) & 4.0 (20) & 7.2 ($\infty$) \\ 
& \multirow{2}{*}{m$_X$ = 700 GeV} & AUC & 0.51 & 0.71 & 0.77 & 0.87 & 0.86 & 0.91 & 0.99 \\ & & Max SIC (new $\sigma$) & 1.0 (0.0) & 1.8 (0.9) & 6.1 (6.1) & 7.6 (15) & 6.8 (20) & 5.5 (28) & 12.6 ($\infty$) \\ 
\hline

\multirow{2}{2cm}{Weakly supervised, hadron \shat}
& \multirow{2}{*}{m$_X$ = 350 GeV} & AUC & 0.54 & 0.58 & 0.75 & 0.88 & 0.88 & 0.93 & 0.99 \\ & & Max SIC (new $\sigma$) & 1.0 (0.0) & 1.0 (0.5) & 1.4 (1.4) & 3.3 (6.6) & 4.6 (14) & 6.1 (31) & 7.6 ($\infty$) \\ 
& \multirow{2}{*}{m$_X$ = 700 GeV} & AUC & 0.27 & 0.43 & 0.79 & 0.84 & 0.9 & 0.91 & 0.99  \\ & & Max SIC (new $\sigma$) & 1.0 (0.0) & 1.0 (0.5) & 2.3 (2.3) & 8.6 (17) & 9.9 (30) & 8.0 (40) & 10.5 ($\infty$) \\ 
\hline

\end{tabular}}
\caption{Summary of all considered training scenarios, the resulting area-under-curve (AUC), and maximum value of the significance improvement characteristic (SIC) curve.   A significance of $\infty$ indicates the case of pure signal versus pure background. The "new $\sigma$" refers to the sensitivity of the signal excess after a neural net selection, which is dictated by the max SIC value.\label{tab:resultsummary}}
\end{centering}
\end{table}

%% file: sections/conclusion.tex
\section{Conclusions}
\label{sec:conclusion}

We present results of high- and variable-dimensional anomaly detection applied to the search for new physics at a future \ee~collider. Radiative return events are leveraged to scan \shat~for new resonant particles decaying to hadrons, while remaining agnostic to the mass scale of new physics. The analysis methodology uses different regions of \shat~to obtain training regions with varying signal-to-background ratios, allowing for the construction of a classifier that uses noisy labels for sensitivity to signal characteristics without relying on an input model. Classifiers are trained over \ee~events modeled as particle flow networks, with up to 150 values characterizing each event. 

When training over a simulated dataset with a realistic luminosity, signals with an initial signal-to-background ratio of $\lesssim 1\sigma$ are distinguishable by this classifier, and their sensitivity can be enhanced by over an order of magnitude. A normalization procedure is employed that removes correlation of the result with \shat~and ensures that no spurious signals are found in background-only analyses. Together, these results indicate the strong potential for this method to facilitate new physics searches in future \ee~datasets. Further studies can be used to inform the design of future detectors by understanding the impact of detector acceptance and resolution on the sensitivity to generic new physics, as well as utilizing full detector simulation. The examination of different input features or even higher input dimensionality could also broaden application of this method beyond hadronically decaying resonant new physics to an even more generic search. These avenues are both closely linked to the development of a robust and exciting physics program at a next-generation \ee~collider experiment. 

%% file: sections/app_pfnvars.tex
\section{Particle Flow Network Variable Distributions}
\label{app:pfnvars}

\begin{figure}
\begin{centering}
\includegraphics[scale=0.3]{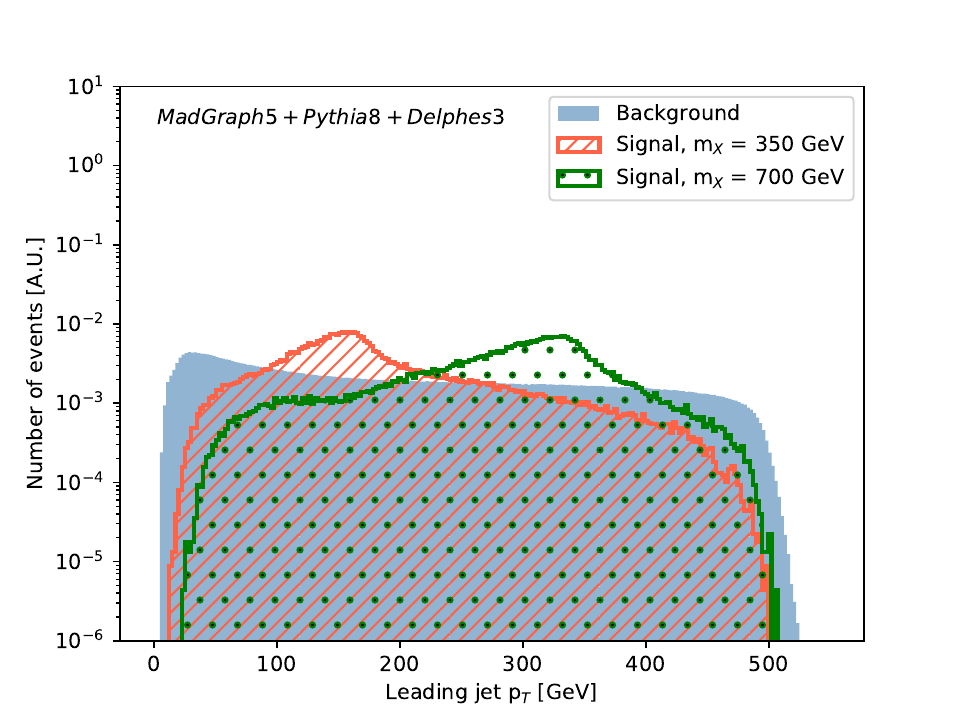}
\includegraphics[scale=0.3]{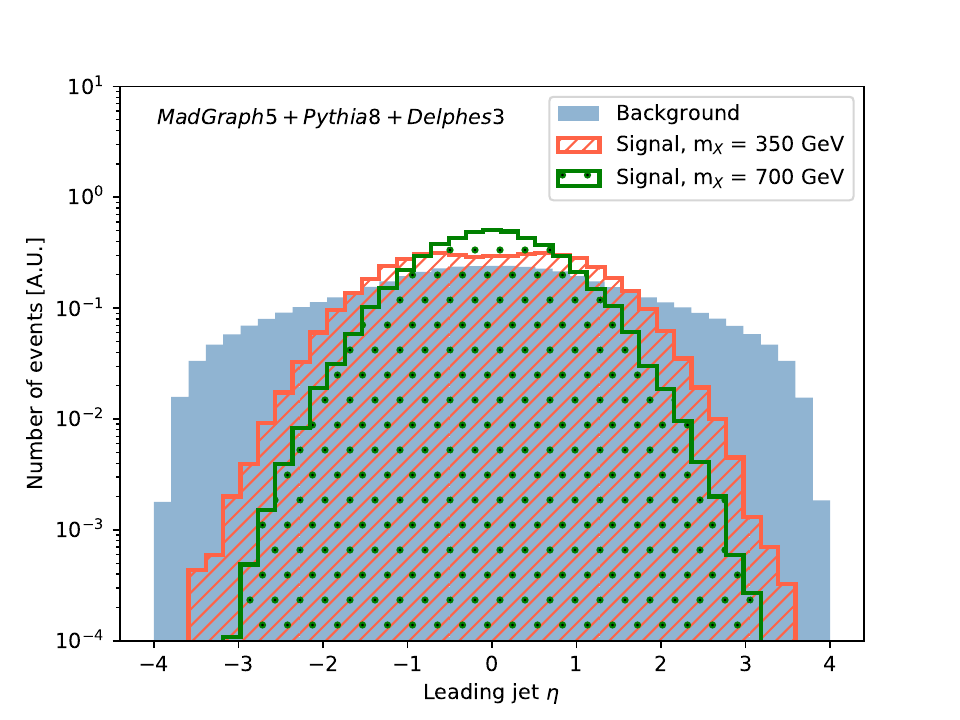}
\includegraphics[scale=0.3]{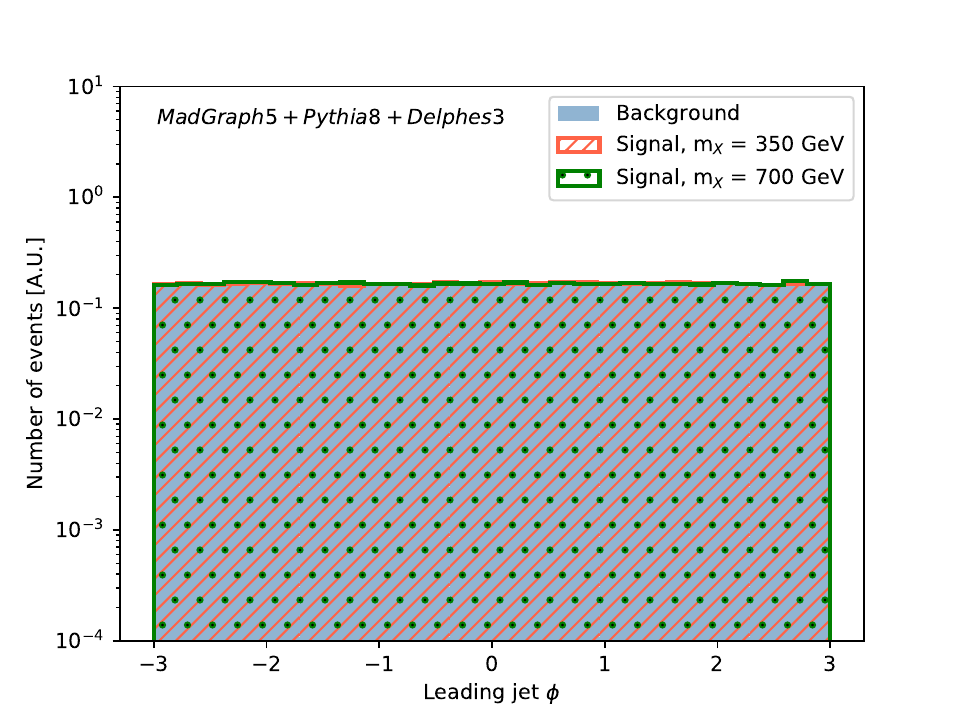}
\includegraphics[scale=0.3]{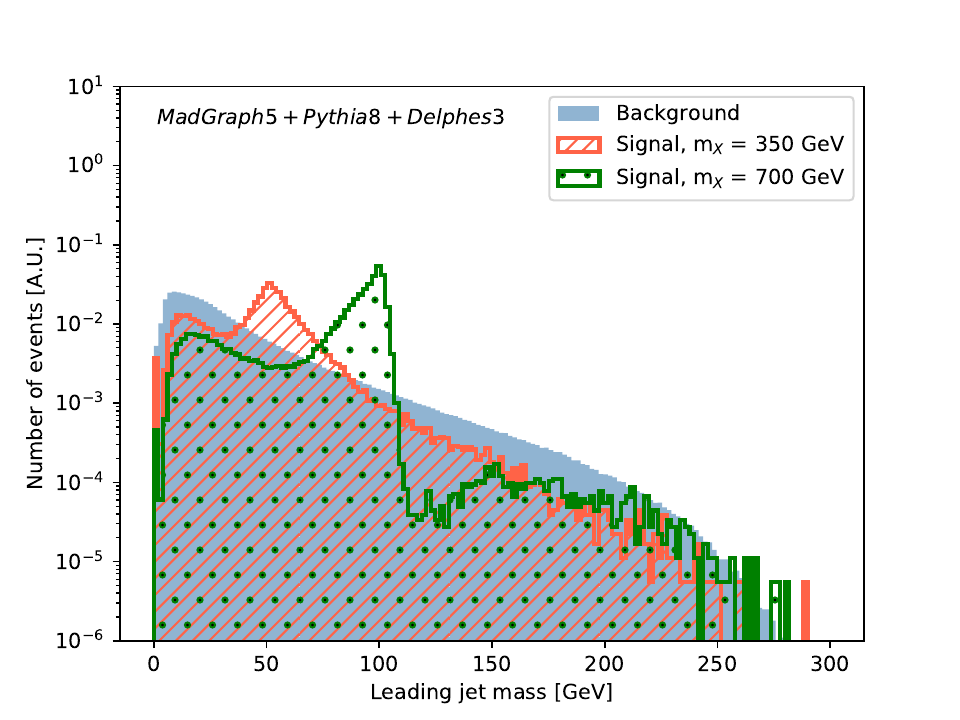}
\includegraphics[scale=0.3]{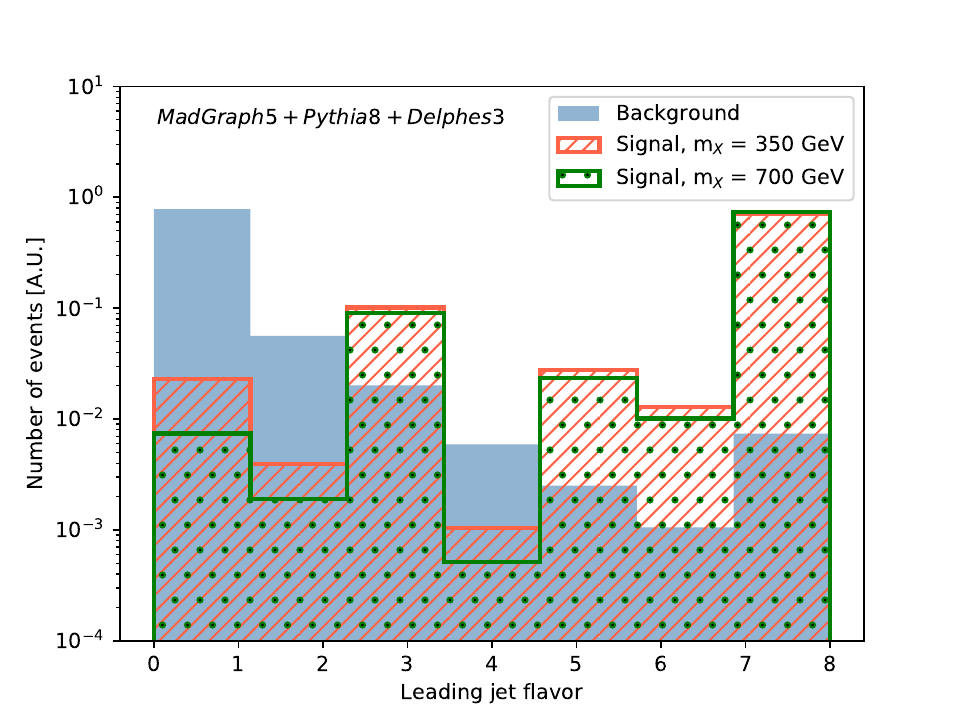}
\includegraphics[scale=0.3]{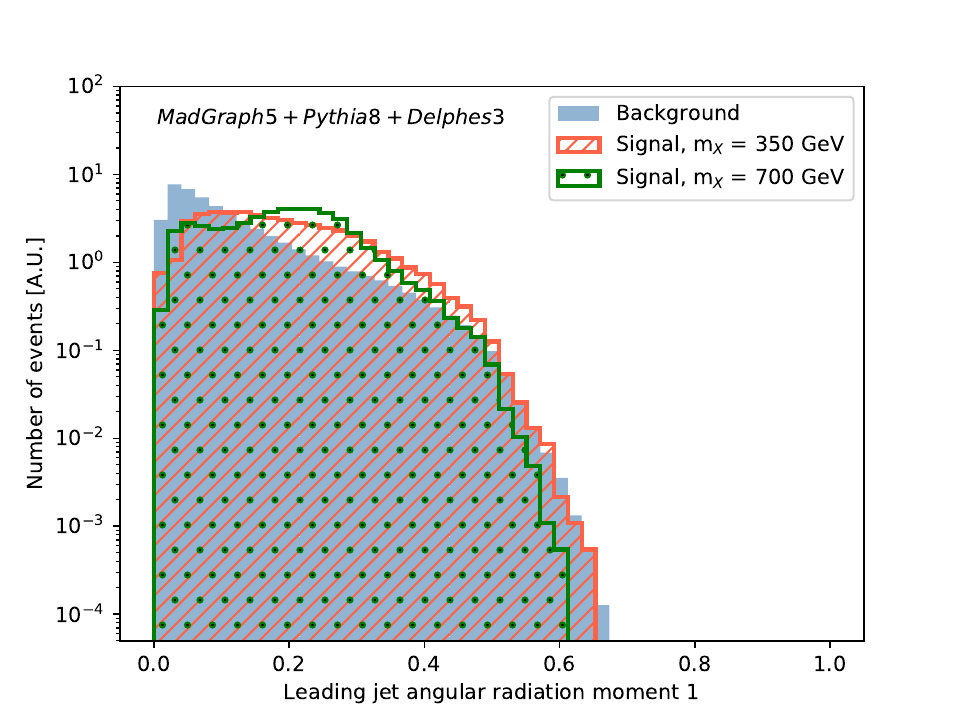}
\includegraphics[scale=0.3]{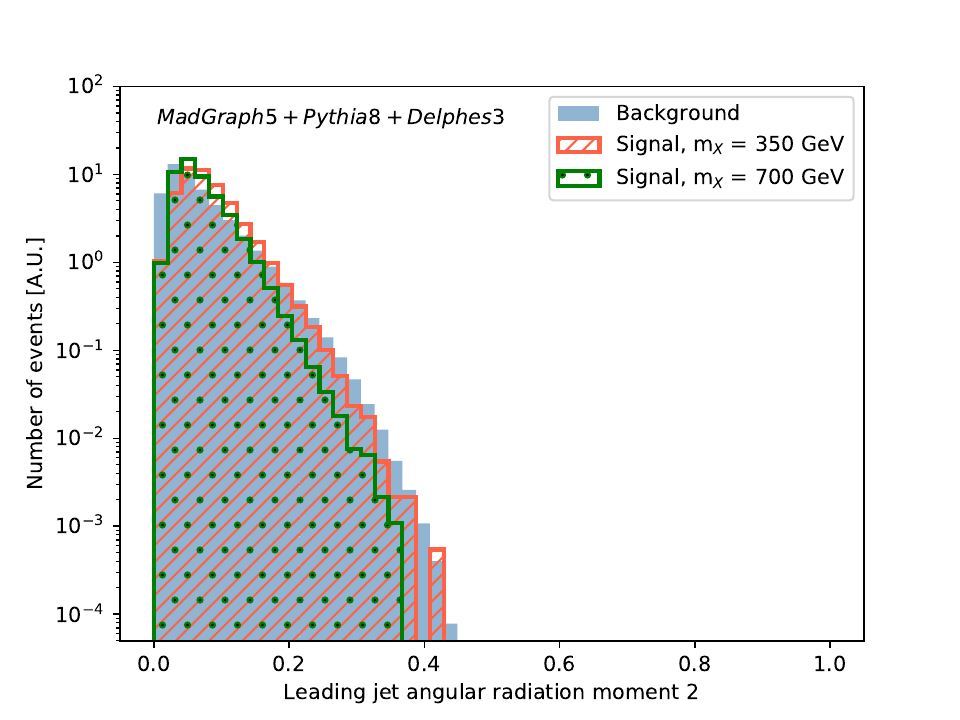}
\includegraphics[scale=0.3]{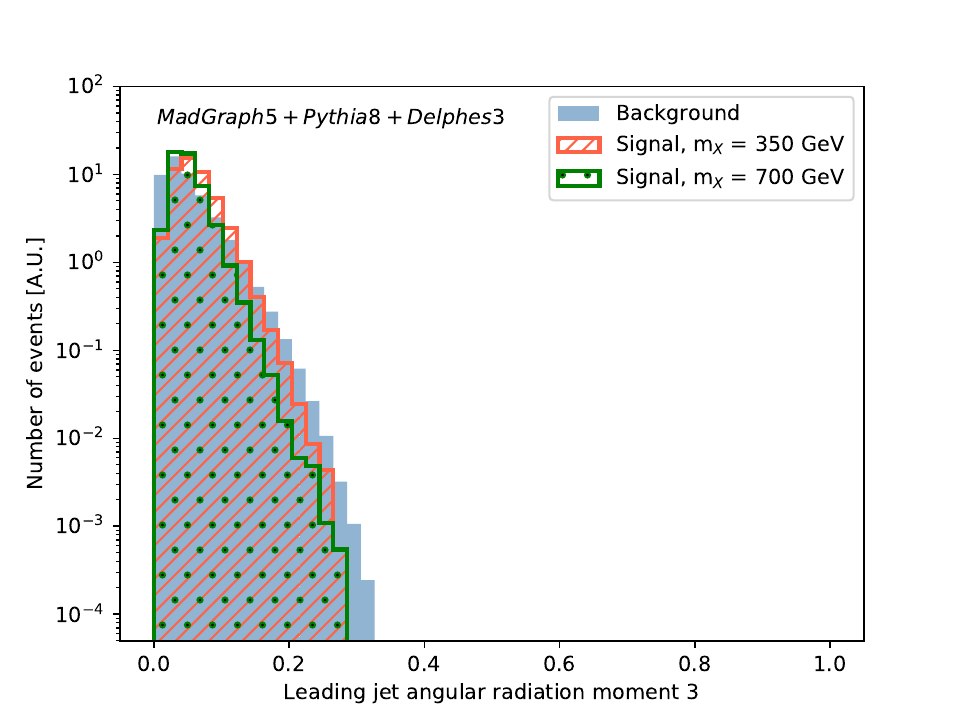}
\includegraphics[scale=0.3]{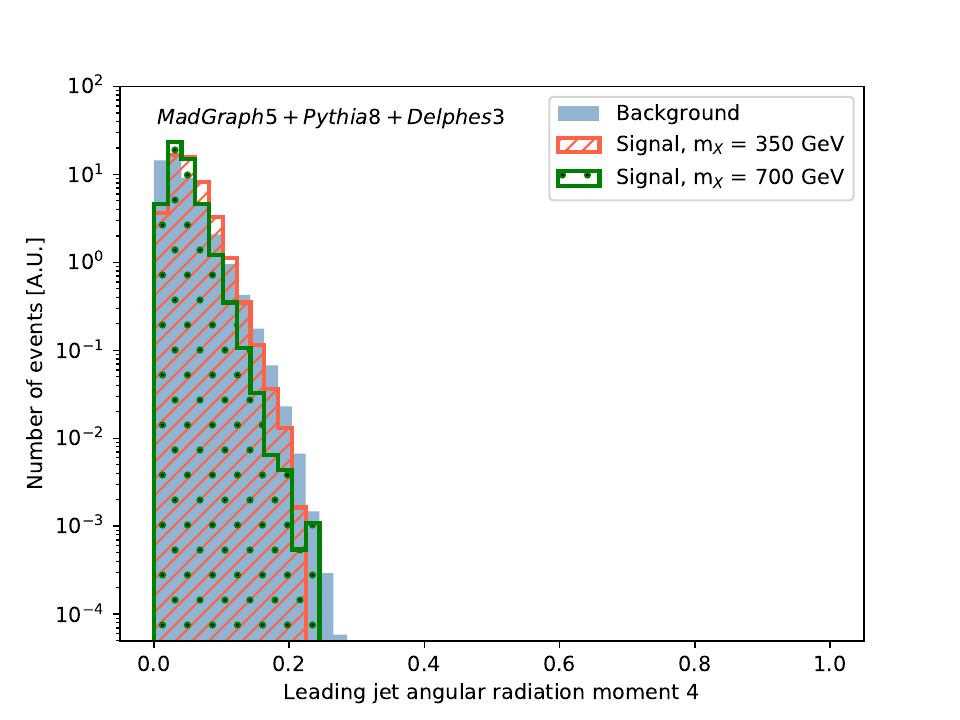}
\includegraphics[scale=0.3]{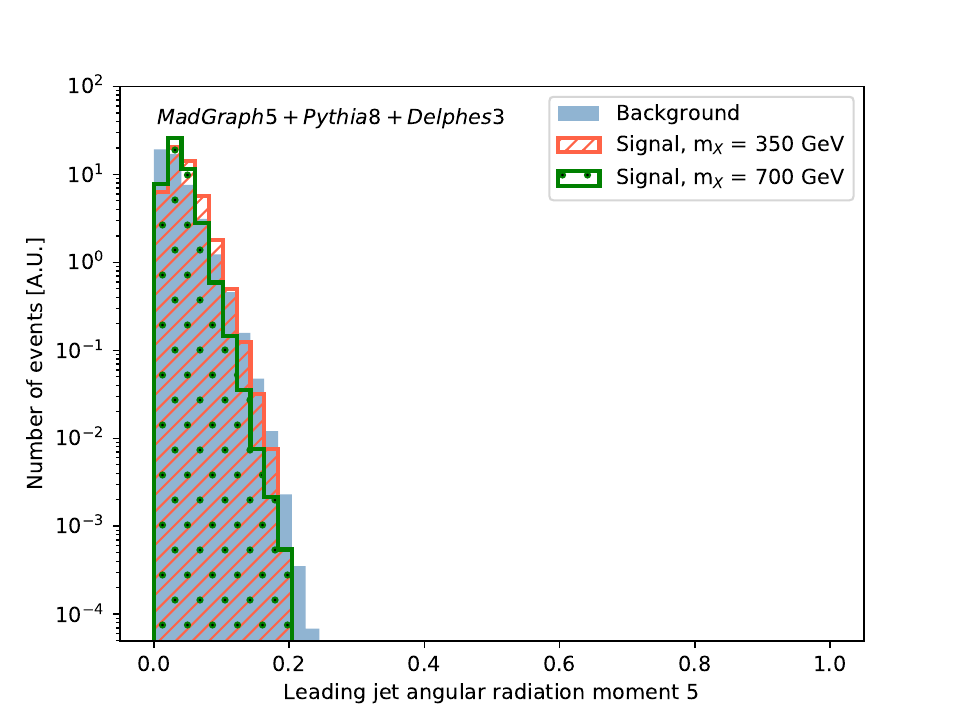}
\caption{Leading jet input PFN variables.\label{fig:pfn_inputs_leading}}
\end{centering}
\end{figure}

\begin{figure}
\begin{centering}
\includegraphics[scale=0.3]{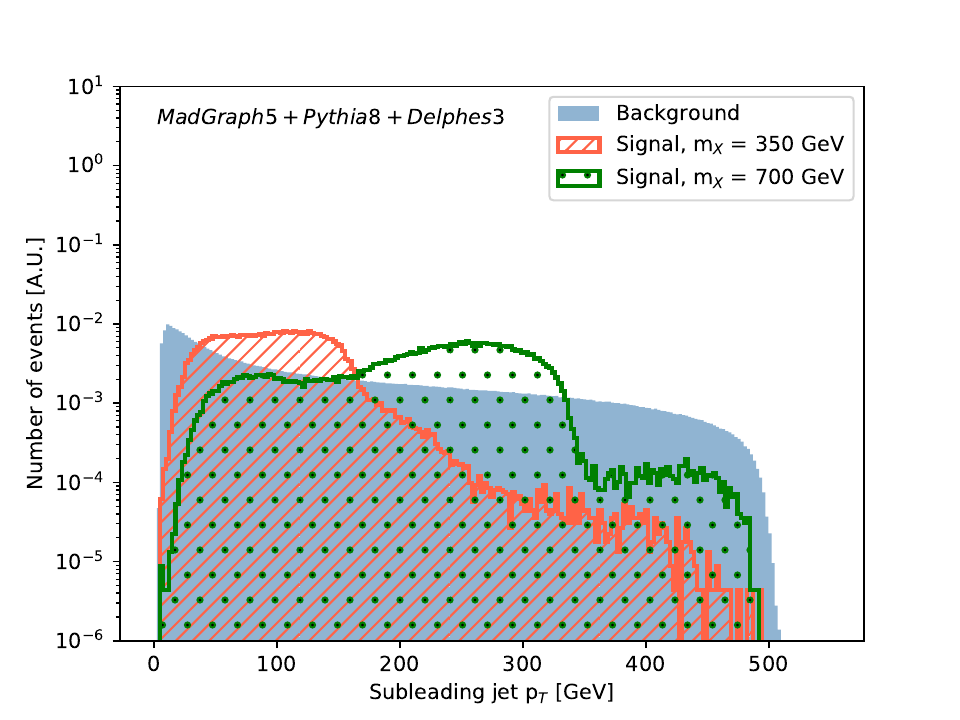}
\includegraphics[scale=0.3]{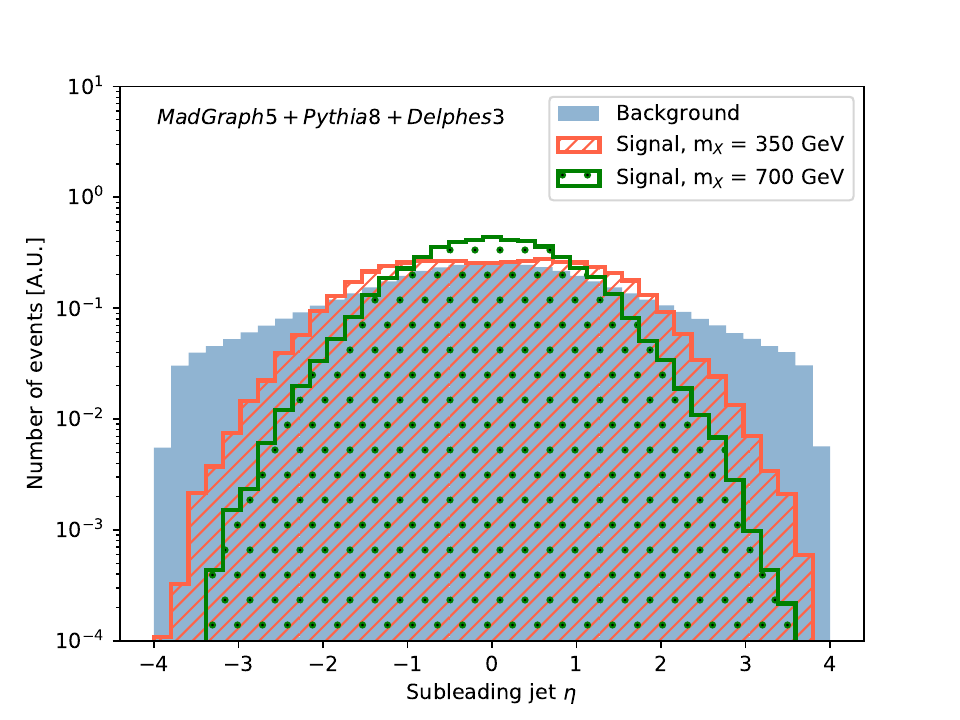}
\includegraphics[scale=0.3]{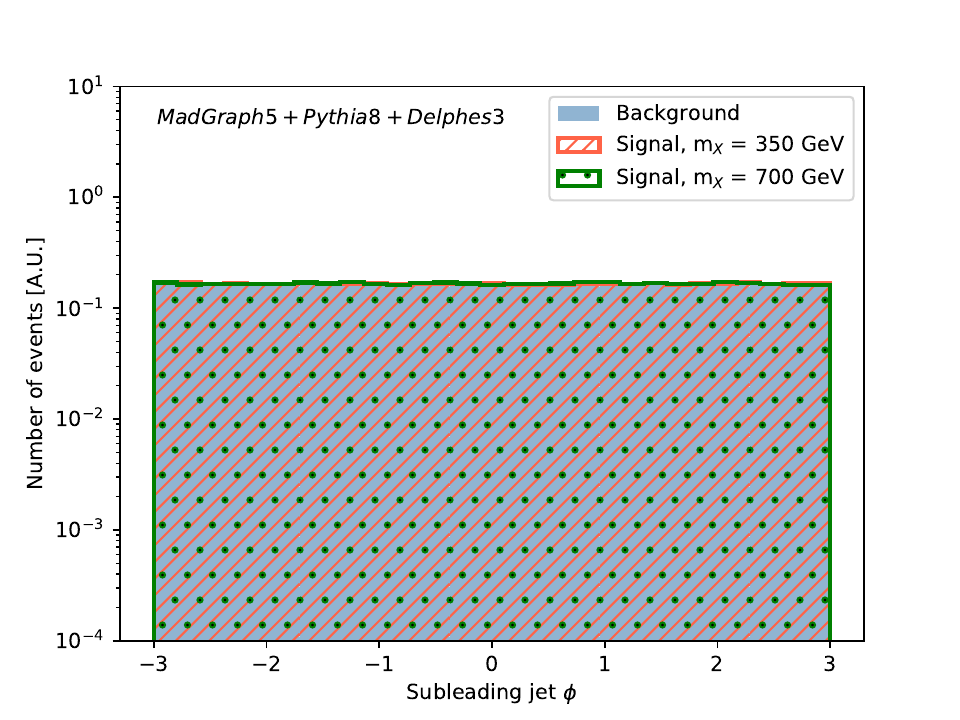}
\includegraphics[scale=0.3]{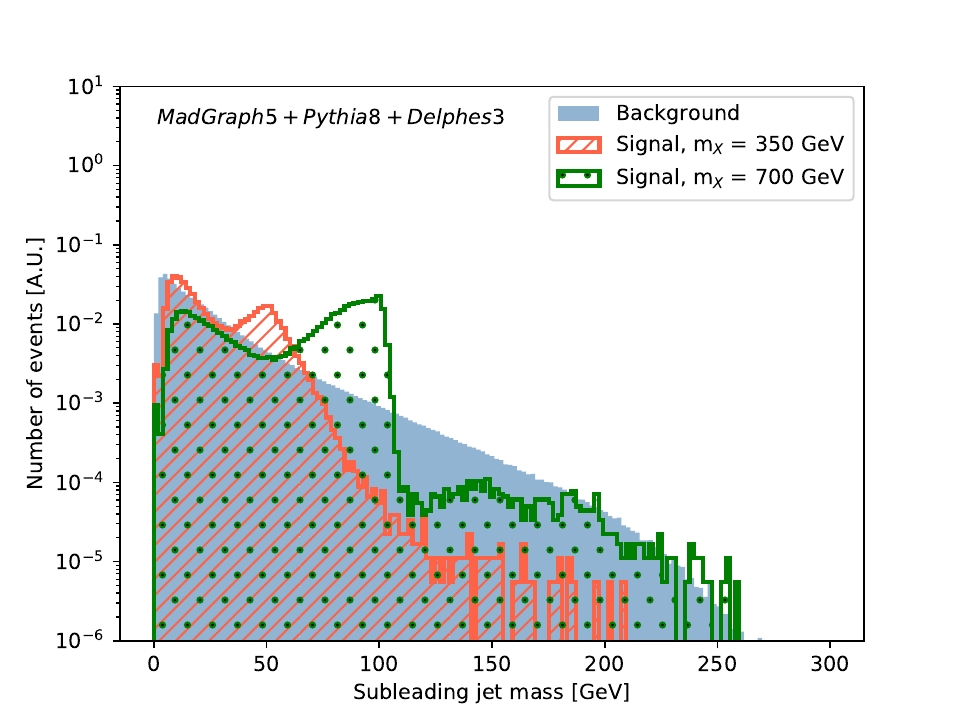}
\includegraphics[scale=0.3]{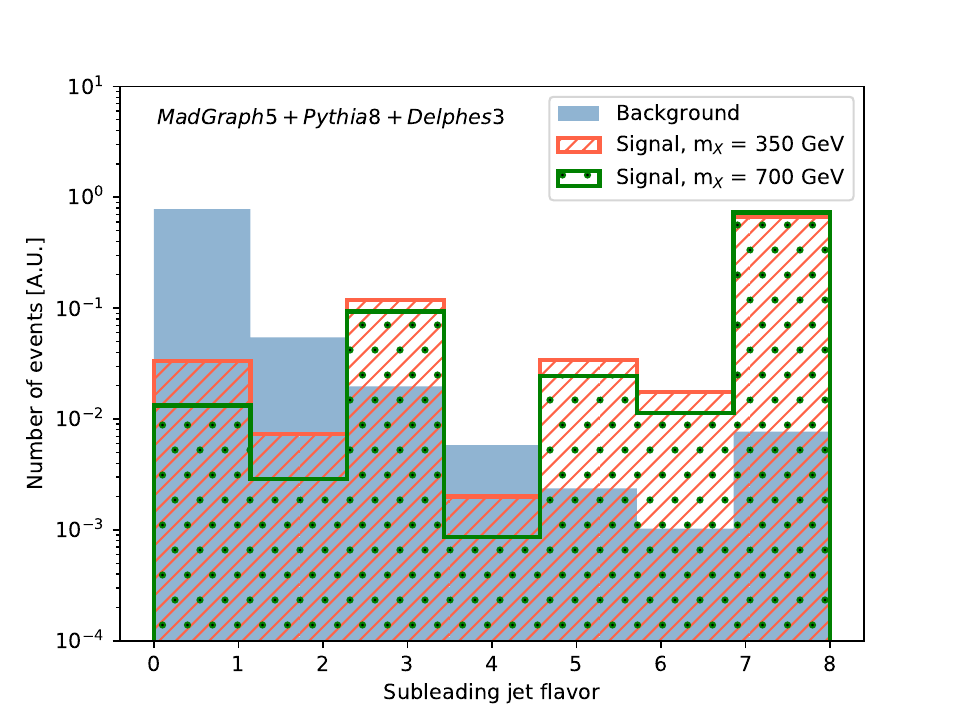}
\includegraphics[scale=0.3]{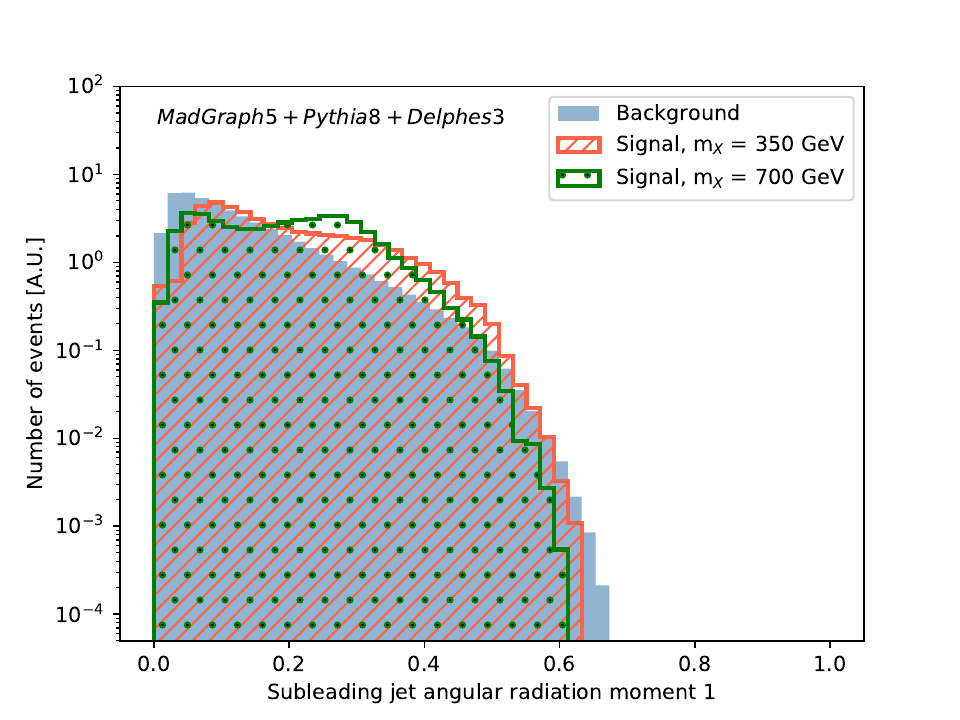}
\includegraphics[scale=0.3]{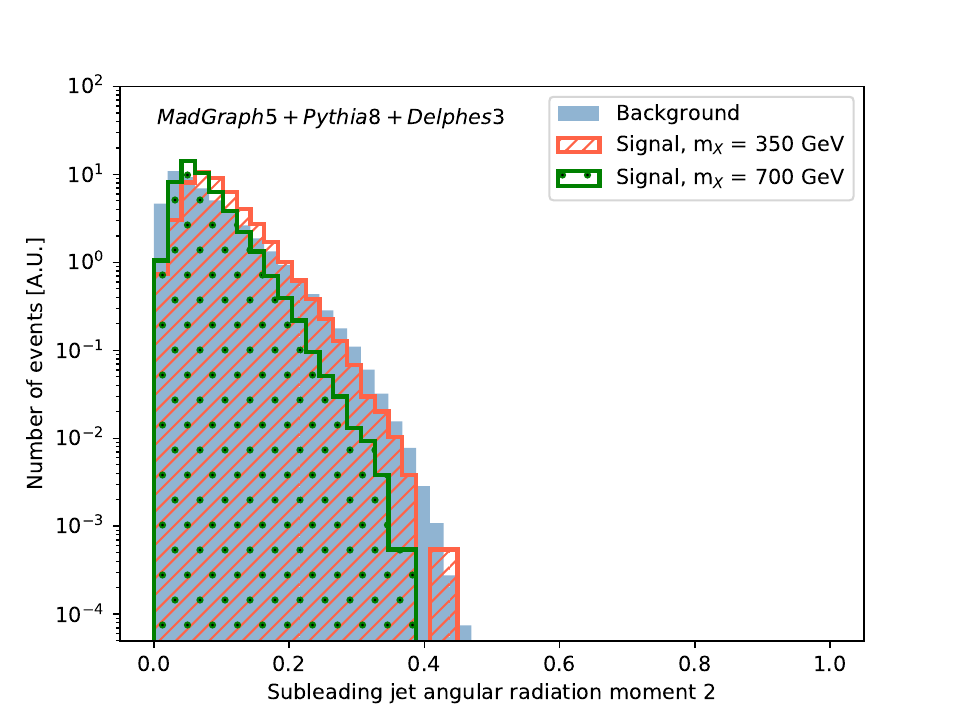}
\includegraphics[scale=0.3]{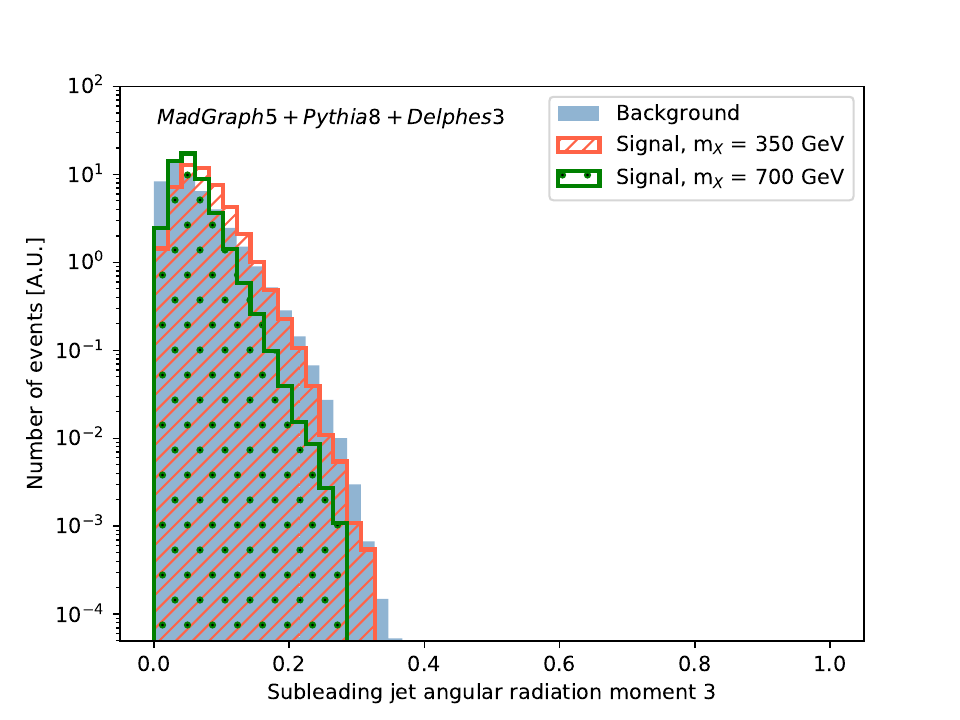}
\includegraphics[scale=0.3]{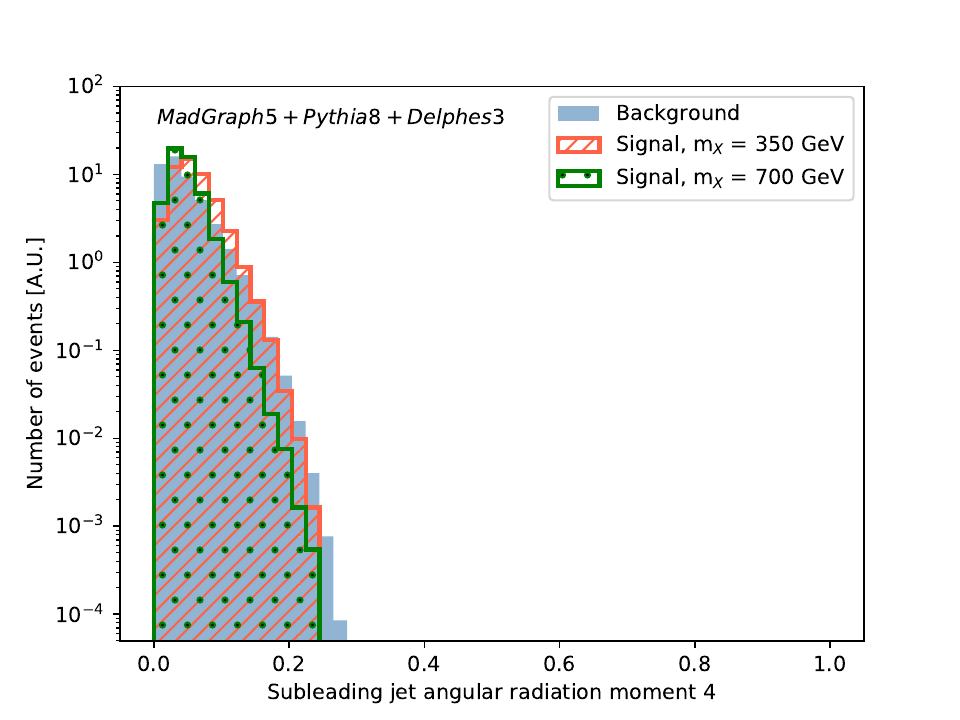}
\includegraphics[scale=0.3]{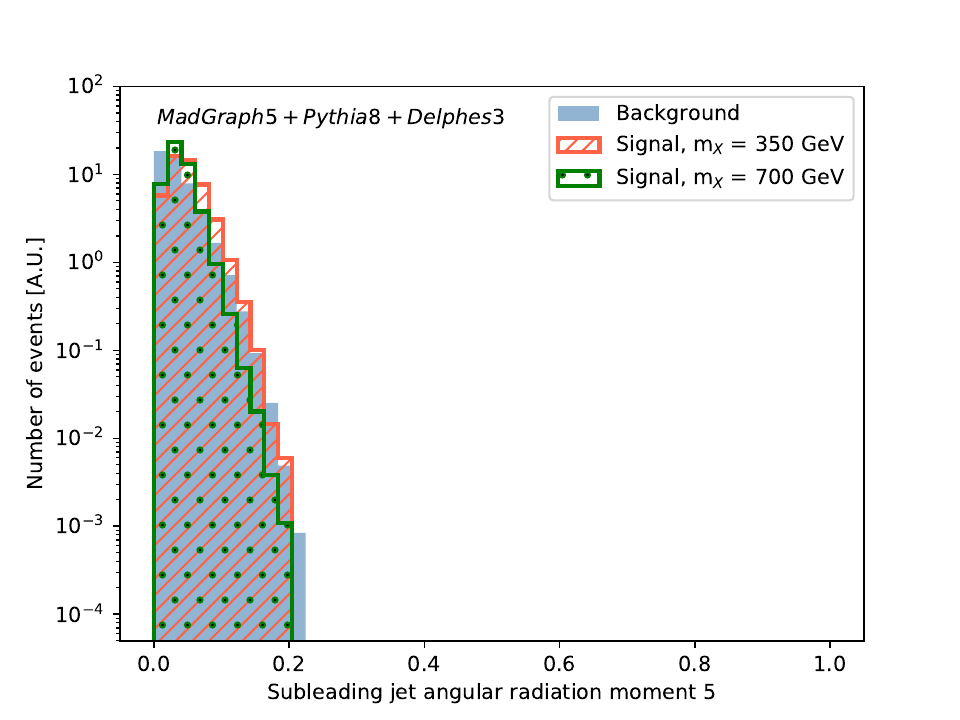}
\caption{Subleading jet input PFN variables.\label{fig:pfn_inputs_subleading}}
\end{centering}
\end{figure}

Figures~\ref{fig:pfn_inputs_leading} and ~\ref{fig:pfn_inputs_subleading} show example distributions of the PFN variables used in training for both signals and the background. Only the leading and subleading jet inputs are shown, though training is done over the same variables for up to 15 jets per event.

%% file: sections/app_eventlevelvars.tex
\section{Event Level Variable Distributions}
\label{app:eventlevelvars}

The following 15 variables were used as input to the event-level DNN approach.
\begin{itemize}
\item Leading and subleading large-R jet masses
\item Leading and subleading large-R jet transverse momenta
\item Leading photon $p_T$
\item Measured $X$ $p_T$, defined by the two leading jets
\item Number of particles per event
\item Number of jets per event
\item Ratio of leading jet $p_T$ to leading photon $p_T$
\item Ratio of measured $X$ $p_T$ to leading photon $p_T$
\item ln$y_{23}$, calculated with all reconstructed jets in the event~\cite{lny23_1,lny23_2}. 
\item Aplanarity, sphericity, and transverse sphericity~\cite{eventshape_opal,eventshape_atlas}
\item Total jet mass~\cite{totaljetmass}.
\end{itemize}

\begin{figure}
\begin{centering}
\includegraphics[scale=0.3]{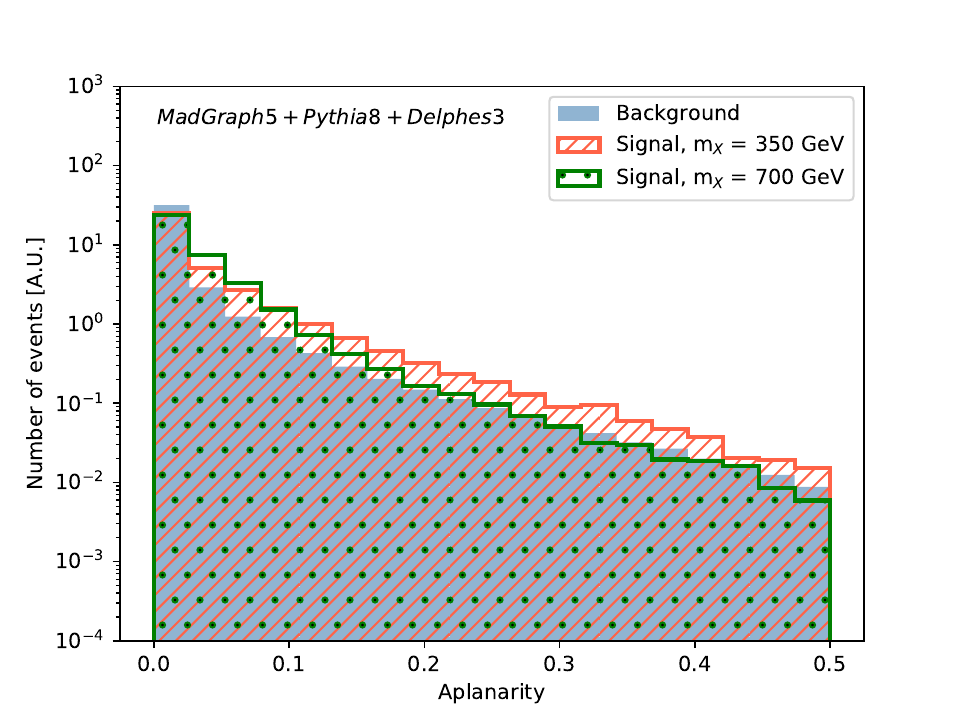}
\includegraphics[scale=0.3]{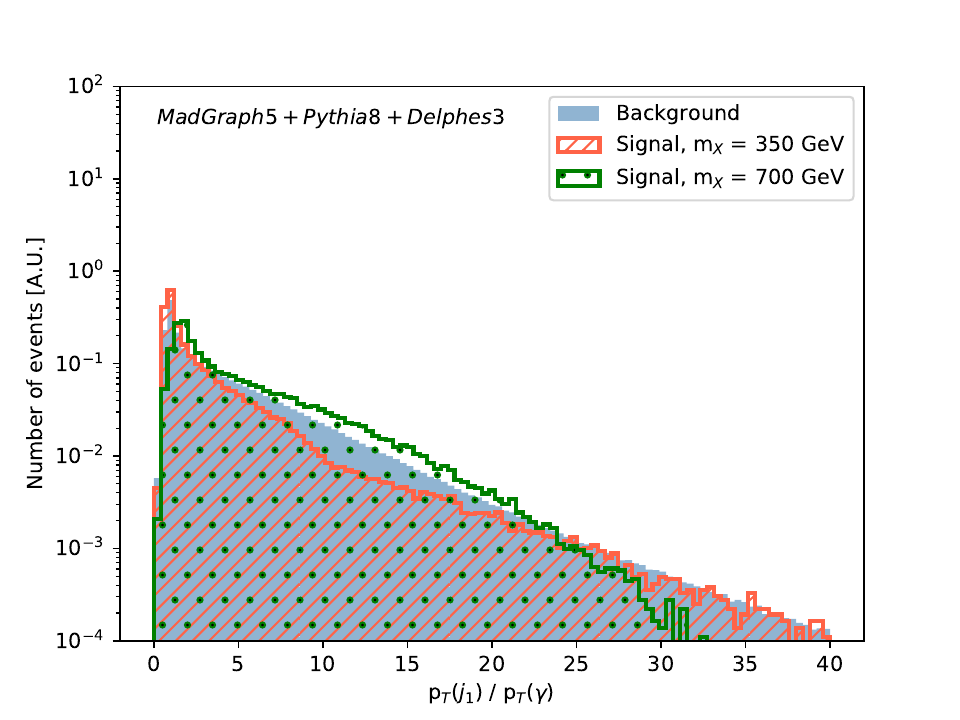}
\includegraphics[scale=0.3]{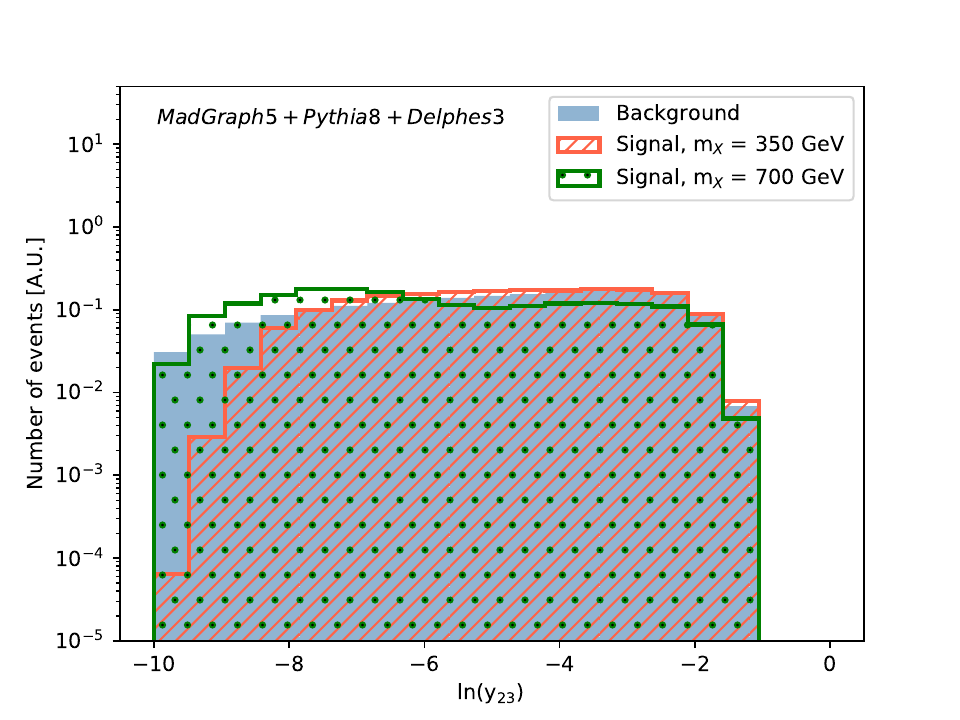}
\includegraphics[scale=0.3]{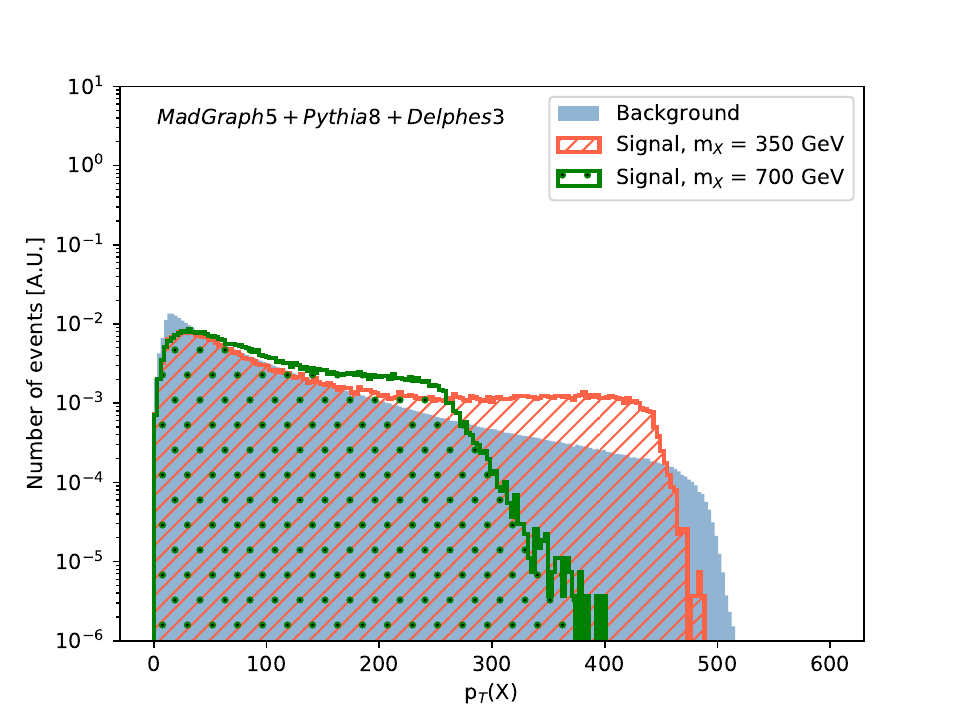}
\includegraphics[scale=0.3]{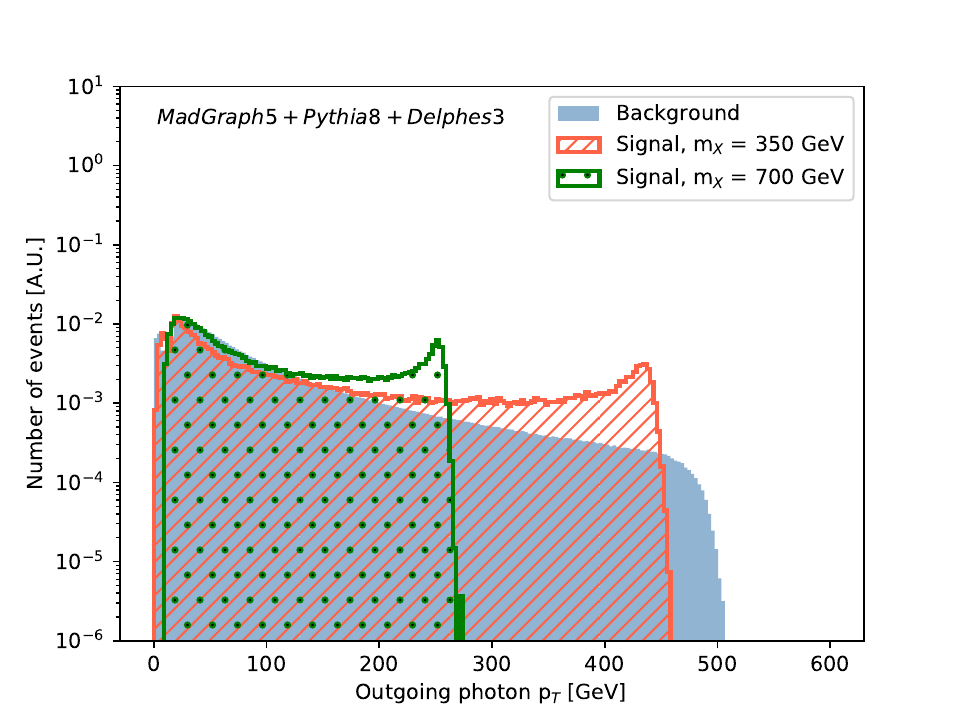}
\includegraphics[scale=0.3]{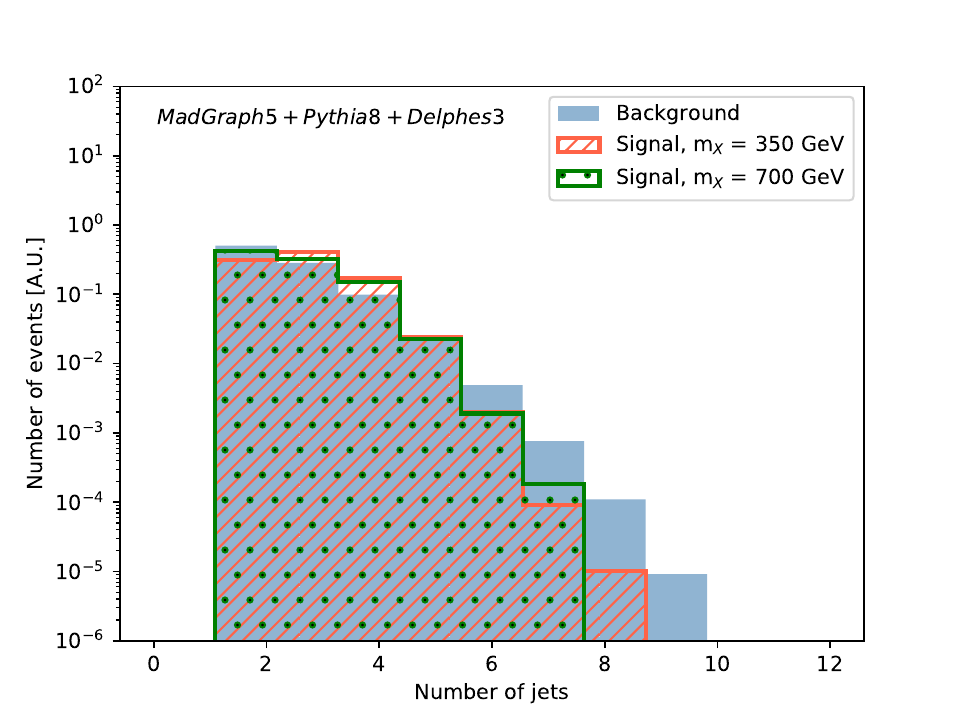}
\includegraphics[scale=0.3]{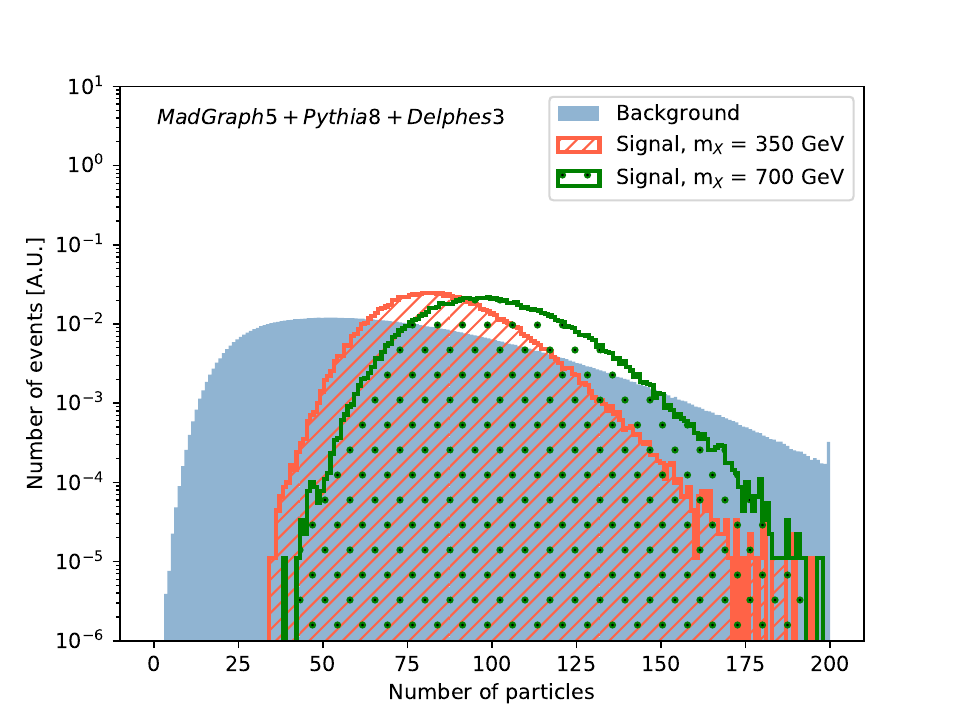}
\includegraphics[scale=0.3]{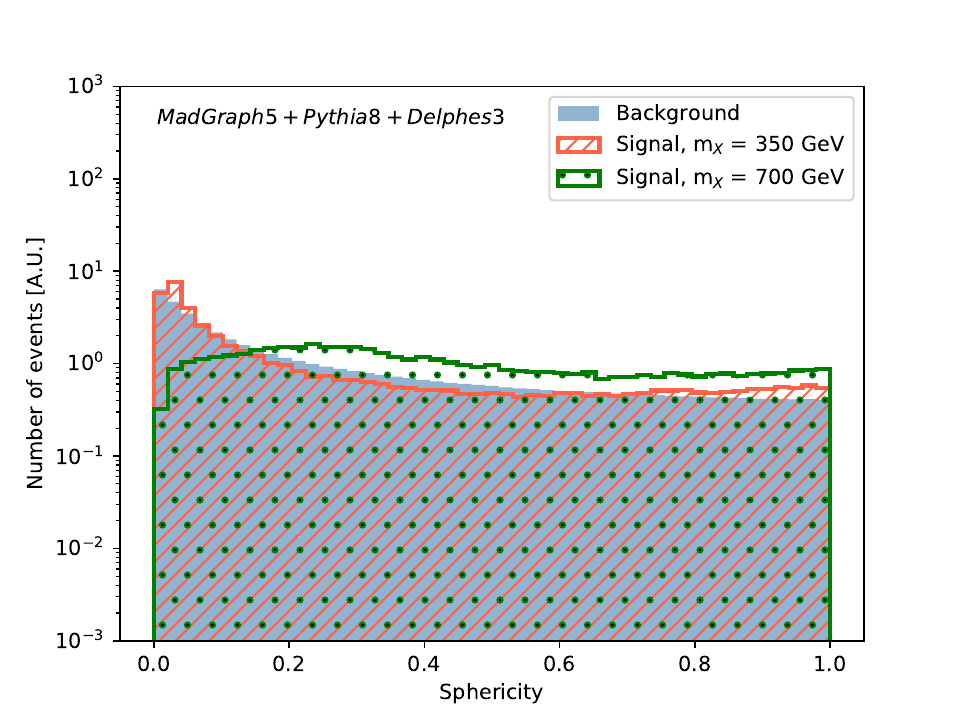}
\includegraphics[scale=0.3]{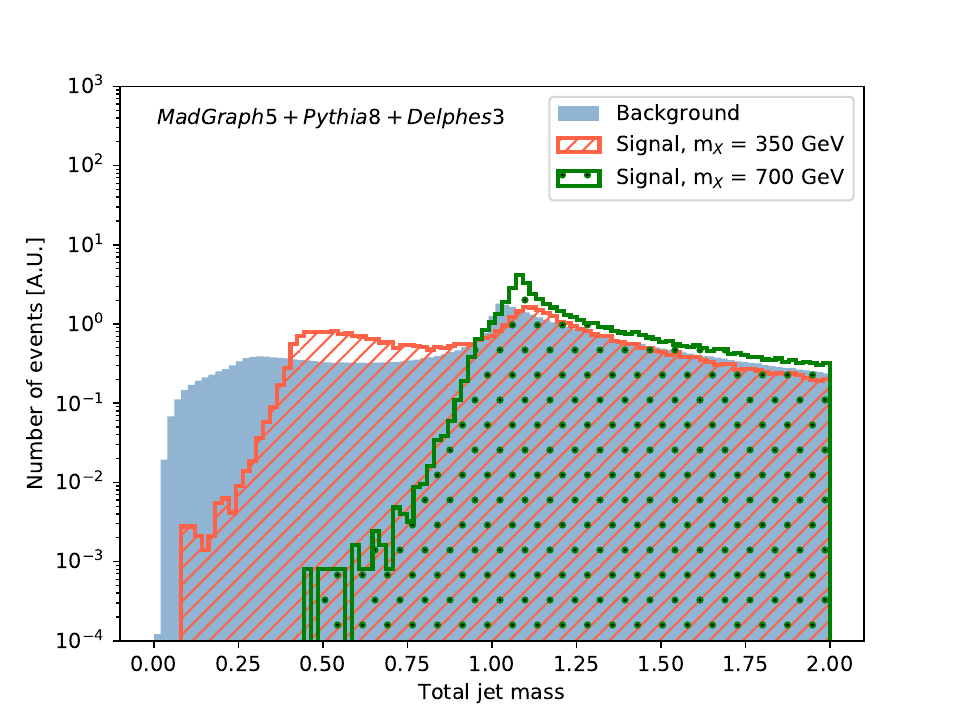}
\includegraphics[scale=0.3]{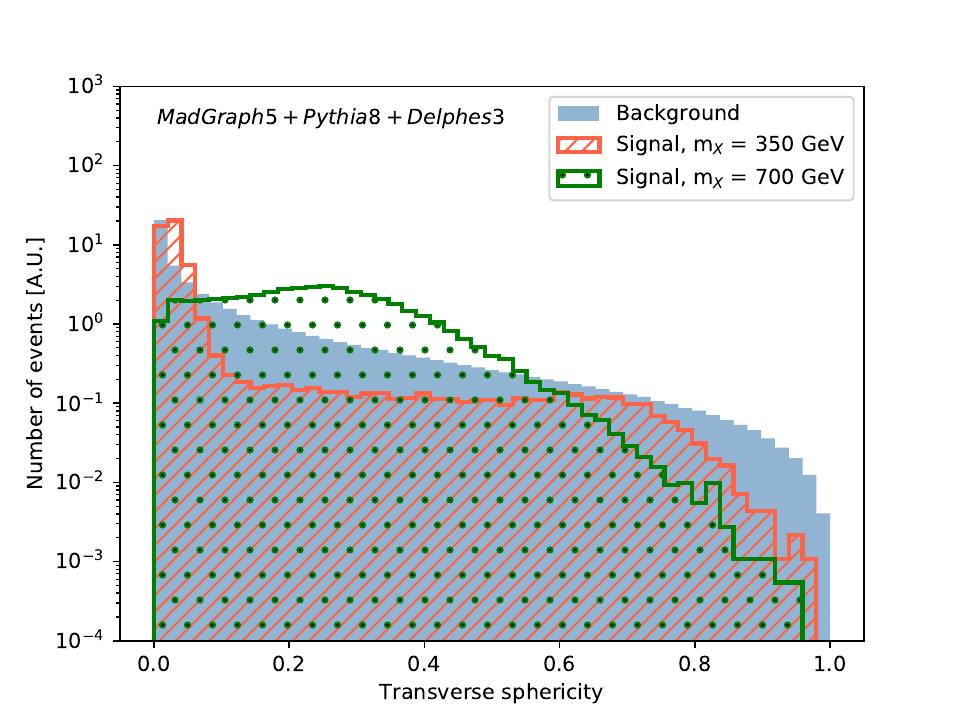}
\includegraphics[scale=0.3]{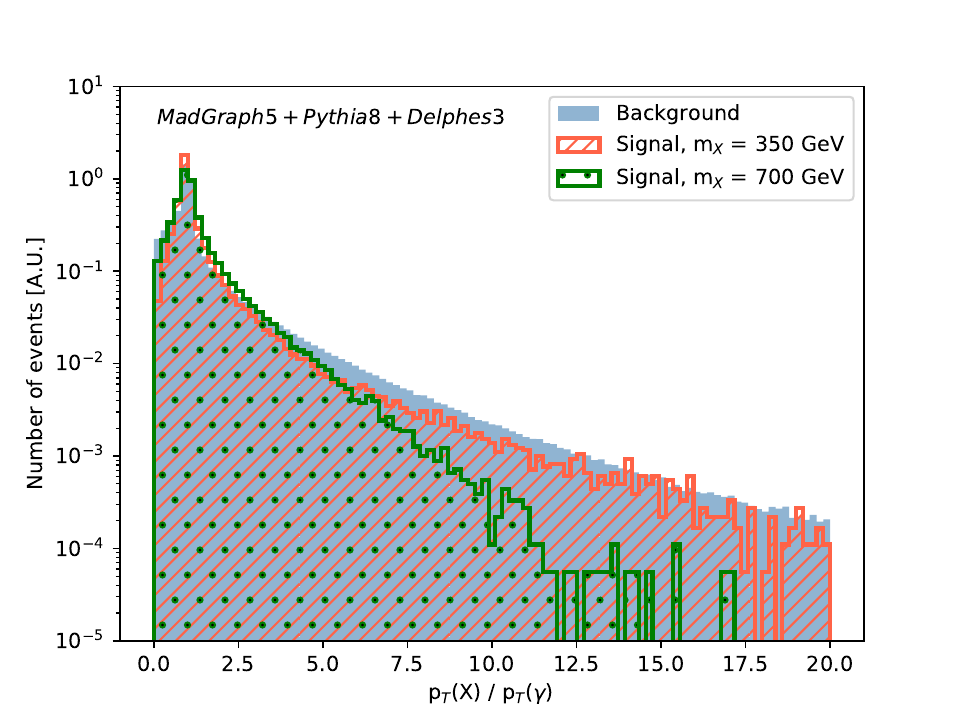}
\caption{Event-level variables used in training, overlaying background and both signal mass hypothesis distributions.\label{fig:eventlevel_inputs}}
\end{centering}
\end{figure}

Figure~\ref{fig:eventlevel_inputs} show the distributions of the event-level variables used in training for the background, $m_X = 350$~GeV signal, and $m_X = 700$~GeV signal. The two leading jet masses and \pt~distributions are also used in the PFN training, and can be seen in Appendix~\ref{app:pfnvars}.

%% file: ilcanomalies.bbl
\providecommand{\href}[2]{#2}\begingroup\raggedright\begin{thebibliography}{10}

\bibitem{PhysRev.126.1858}
J.~Button, G.~R. Kalbfleisch, G.~R. Lynch, B.~C. Magli\ifmmode~\acute{c}\else
  \'{c}\fi{}, A.~H. Rosenfeld, and M.~L. Stevenson, {\it Pion-pion interaction
  in the reaction
  $\overline{p}+p\ensuremath{\rightarrow}2{\ensuremath{\pi}}^{+}+2{\ensuremath{\pi}}^{\ensuremath{-}}+n{\ensuremath{\pi}}^{0}$},
  {\em Phys. Rev.} {\bf 126} (Jun, 1962) 1858--1863.

\bibitem{atlas_higgs}
{\relax ATLAS Collaboration}, {\it Observation of a new particle in the search
  for the {Standard} {Model} {Higgs} boson with the {ATLAS} detector at the
  {LHC}},  {\em Physics Letters B} {\bf 716} (Sep, 2012) 1--29.

\bibitem{cms_higgs}
{\relax CMS Collaboration}, {\it {Observation of a new boson at a mass of 125
  {GeV} with the {CMS} experiment at the {LHC}}},  {\em Physics Letters B} {\bf
  716} (Sep, 2012) 30--61.

\bibitem{DAgnolo:2018cun}
R.~T. D'Agnolo and A.~Wulzer, {\it {Learning New Physics from a Machine}},
  {\em Phys. Rev.} {\bf D99} (2019), no.~1 015014,
  [\href{http://arxiv.org/abs/1806.02350}{{\tt arXiv:1806.02350}}].

\bibitem{Collins:2018epr}
J.~H. Collins, K.~Howe, and B.~Nachman, {\it {Anomaly Detection for Resonant
  New Physics with Machine Learning}},  {\em Phys. Rev. Lett.} {\bf 121}
  (2018), no.~24 241803, [\href{http://arxiv.org/abs/1805.02664}{{\tt
  arXiv:1805.02664}}].

\bibitem{Collins:2019jip}
J.~H. Collins, K.~Howe, and B.~Nachman, {\it {Extending the search for new
  resonances with machine learning}},  {\em Phys. Rev.} {\bf D99} (2019), no.~1
  014038, [\href{http://arxiv.org/abs/1902.02634}{{\tt arXiv:1902.02634}}].

\bibitem{DAgnolo:2019vbw}
R.~T. D'Agnolo, G.~Grosso, M.~Pierini, A.~Wulzer, and M.~Zanetti, {\it
  {Learning Multivariate New Physics}},
  \href{http://arxiv.org/abs/1912.12155}{{\tt arXiv:1912.12155}}.

\bibitem{Farina:2018fyg}
M.~Farina, Y.~Nakai, and D.~Shih, {\it {Searching for New Physics with Deep
  Autoencoders}},  \href{http://arxiv.org/abs/1808.08992}{{\tt
  arXiv:1808.08992}}.

\bibitem{Heimel:2018mkt}
T.~Heimel, G.~Kasieczka, T.~Plehn, and J.~M. Thompson, {\it {QCD or What?}},
  {\em SciPost Phys.} {\bf 6} (2019), no.~3 030,
  [\href{http://arxiv.org/abs/1808.08979}{{\tt arXiv:1808.08979}}].

\bibitem{Roy:2019jae}
T.~S. Roy and A.~H. Vijay, {\it {A robust anomaly finder based on
  autoencoder}},  \href{http://arxiv.org/abs/1903.02032}{{\tt
  arXiv:1903.02032}}.

\bibitem{Cerri:2018anq}
O.~Cerri, T.~Q. Nguyen, M.~Pierini, M.~Spiropulu, and J.-R. Vlimant, {\it
  {Variational Autoencoders for New Physics Mining at the Large Hadron
  Collider}},  {\em JHEP} {\bf 05} (2019) 036,
  [\href{http://arxiv.org/abs/1811.10276}{{\tt arXiv:1811.10276}}].

\bibitem{Blance:2019ibf}
A.~Blance, M.~Spannowsky, and P.~Waite, {\it {Adversarially-trained
  autoencoders for robust unsupervised new physics searches}},  {\em JHEP} {\bf
  10} (2019) 047, [\href{http://arxiv.org/abs/1905.10384}{{\tt
  arXiv:1905.10384}}].

\bibitem{Hajer:2018kqm}
J.~Hajer, Y.-Y. Li, T.~Liu, and H.~Wang, {\it {Novelty Detection Meets Collider
  Physics}},  \href{http://arxiv.org/abs/1807.10261}{{\tt arXiv:1807.10261}}.

\bibitem{DeSimone:2018efk}
A.~De~Simone and T.~Jacques, {\it {Guiding New Physics Searches with
  Unsupervised Learning}},  {\em Eur. Phys. J.} {\bf C79} (2019), no.~4 289,
  [\href{http://arxiv.org/abs/1807.06038}{{\tt arXiv:1807.06038}}].

\bibitem{1809.02977}
G.~M. Alessandro~Casa, {\it {Nonparametric semisupervised classification for
  signal detection in high energy physics}},
  \href{http://arxiv.org/abs/1809.02977}{{\tt arXiv:1809.02977}}.

\bibitem{Dillon:2019cqt}
B.~M. Dillon, D.~A. Faroughy, and J.~F. Kamenik, {\it {Uncovering latent jet
  substructure}},  {\em Phys. Rev.} {\bf D100} (2019), no.~5 056002,
  [\href{http://arxiv.org/abs/1904.04200}{{\tt arXiv:1904.04200}}].

\bibitem{Andreassen:2020nkr}
A.~Andreassen, B.~Nachman, and D.~Shih, {\it {Simulation Assisted
  Likelihood-free Anomaly Detection}},  {\em Phys. Rev. D} {\bf 101} (2020),
  no.~9 095004, [\href{http://arxiv.org/abs/2001.05001}{{\tt
  arXiv:2001.05001}}].

\bibitem{Nachman:2020lpy}
B.~Nachman and D.~Shih, {\it {Anomaly Detection with Density Estimation}},
  {\em Phys. Rev. D} {\bf 101} (2020) 075042,
  [\href{http://arxiv.org/abs/2001.04990}{{\tt arXiv:2001.04990}}].

\bibitem{Aguilar-Saavedra:2017rzt}
J.~A. Aguilar-Saavedra, J.~H. Collins, and R.~K. Mishra, {\it {A generic
  anti-QCD jet tagger}},  {\em JHEP} {\bf 11} (2017) 163,
  [\href{http://arxiv.org/abs/1709.01087}{{\tt arXiv:1709.01087}}].

\bibitem{Romao:2019dvs}
M.~Romão~Crispim, N.~Castro, R.~Pedro, and T.~Vale, {\it {Transferability of
  Deep Learning Models in Searches for New Physics at Colliders}},  {\em Phys.\
  Rev.\ D} {\bf 101} (2020), no.~3 035042,
  [\href{http://arxiv.org/abs/1912.04220}{{\tt arXiv:1912.04220}}].

\bibitem{Romao:2020ojy}
M.~C. Romao, N.~Castro, J.~Milhano, R.~Pedro, and T.~Vale, {\it {Use of a
  Generalized Energy Mover's Distance in the Search for Rare Phenomena at
  Colliders}},  \href{http://arxiv.org/abs/2004.09360}{{\tt arXiv:2004.09360}}.

\bibitem{knapp2020adversarially}
O.~Knapp, G.~Dissertori, O.~Cerri, T.~Q. Nguyen, J.-R. Vlimant, and M.~Pierini,
  {\it {Adversarially Learned Anomaly Detection on CMS Open Data:
  re-discovering the top quark}},  \href{http://arxiv.org/abs/2005.01598}{{\tt
  arXiv:2005.01598}}.

\bibitem{collaboration2020dijet}
{ATLAS Collaboration}, {\it {Dijet resonance search with weak supervision using
  13 TeV pp collisions in the ATLAS detector}},
  \href{http://arxiv.org/abs/2005.02983}{{\tt arXiv:2005.02983}}.

\bibitem{1797846}
B.~M. Dillon, D.~A. Faroughy, J.~F. Kamenik, and M.~Szewc, {\it {Learning the
  latent structure of collider events}},
  \href{http://arxiv.org/abs/2005.12319}{{\tt arXiv:2005.12319}}.

\bibitem{1800445}
M.~C. Romao, N.~Castro, and R.~Pedro, {\it {Finding New Physics without
  learning about it: Anomaly Detection as a tool for Searches at Colliders}},
  \href{http://arxiv.org/abs/2006.05432}{{\tt arXiv:2006.05432}}.

\bibitem{Amram:2020ykb}
O.~Amram and C.~M. Suarez, {\it {Tag N' Train: A Technique to Train Improved
  Classifiers on Unlabeled Data}},  \href{http://arxiv.org/abs/2002.12376}{{\tt
  arXiv:2002.12376}}.

\bibitem{Cheng:2020dal}
T.~Cheng, J.-F. Arguin, J.~Leissner-Martin, J.~Pilette, and T.~Golling, {\it
  {Variational Autoencoders for Anomalous Jet Tagging}},
  \href{http://arxiv.org/abs/2007.01850}{{\tt arXiv:2007.01850}}.

\bibitem{Khosa:2020qrz}
C.~K. Khosa and V.~Sanz, {\it {Anomaly Awareness}},
  \href{http://arxiv.org/abs/2007.14462}{{\tt arXiv:2007.14462}}.

\bibitem{aguilarsaavedra2020mass}
J.~A. Aguilar-Saavedra, F.~R. Joaquim, and J.~F. Seabra, {\it {Mass Unspecific
  Supervised Tagging (MUST) for boosted jets}},
  \href{http://arxiv.org/abs/2008.12792}{{\tt arXiv:2008.12792}}.

\bibitem{1815227}
K.~Benkendorfer, L.~L. Pottier, and B.~Nachman, {\it {Simulation-Assisted
  Decorrelation for Resonant Anomaly Detection}},
  \href{http://arxiv.org/abs/2009.02205}{{\tt arXiv:2009.02205}}.

\bibitem{pol2020anomaly}
{Adrian Alan Pol and Victor Berger and Gianluca Cerminara and Cecile Germain
  and Maurizio Pierini}, {\it {Anomaly Detection With Conditional Variational
  Autoencoders}},  \href{http://arxiv.org/abs/2010.05531}{{\tt
  arXiv:2010.05531}}.

\bibitem{Mikuni:2020qds}
V.~Mikuni and F.~Canelli, {\it {Unsupervised clustering for collider physics}},
   \href{http://arxiv.org/abs/2010.07106}{{\tt arXiv:2010.07106}}.

\bibitem{vanBeekveld:2020txa}
M.~van Beekveld, S.~Caron, L.~Hendriks, P.~Jackson, A.~Leinweber, S.~Otten,
  R.~Patrick, R.~Ruiz~de Austri, M.~Santoni, and M.~White, {\it {Combining
  outlier analysis algorithms to identify new physics at the LHC}},
  \href{http://arxiv.org/abs/2010.07940}{{\tt arXiv:2010.07940}}.

\bibitem{Park:2020pak}
S.~E. Park, D.~Rankin, S.-M. Udrescu, M.~Yunus, and P.~Harris, {\it {Quasi
  Anomalous Knowledge: Searching for new physics with embedded knowledge}},
  \href{http://arxiv.org/abs/2011.03550}{{\tt arXiv:2011.03550}}.

\bibitem{Faroughy:2020gas}
D.~A. Faroughy, {\it {Uncovering hidden patterns in collider events with
  Bayesian probabilistic models}},  \href{http://arxiv.org/abs/2012.08579}{{\tt
  arXiv:2012.08579}}.

\bibitem{Stein:2020rou}
G.~Stein, U.~Seljak, and B.~Dai, {\it {Unsupervised in-distribution anomaly
  detection of new physics through conditional density estimation}},
  \href{http://arxiv.org/abs/2012.11638}{{\tt arXiv:2012.11638}}.

\bibitem{Kasieczka:2021xcg}
G.~Kasieczka et~al., {\it {The LHC Olympics 2020: A Community Challenge for
  Anomaly Detection in High Energy Physics}},
  \href{http://arxiv.org/abs/2101.08320}{{\tt arXiv:2101.08320}}.

\bibitem{Blance:2021gcs}
A.~Blance and M.~Spannowsky, {\it {Unsupervised Event Classification with
  Graphs on Classical and Photonic Quantum Computers}},
  \href{http://arxiv.org/abs/2103.03897}{{\tt arXiv:2103.03897}}.

\bibitem{Bortolato:2021zic}
B.~Bortolato, B.~M. Dillon, J.~F. Kamenik, and A.~Smolkovi\v{c}, {\it {Bump
  Hunting in Latent Space}},  \href{http://arxiv.org/abs/2103.06595}{{\tt
  arXiv:2103.06595}}.

\bibitem{Collins:2021nxn}
J.~H. Collins, P.~Mart\'\i{}n-Ramiro, B.~Nachman, and D.~Shih, {\it {Comparing
  Weak- and Unsupervised Methods for Resonant Anomaly Detection}},
  \href{http://arxiv.org/abs/2104.02092}{{\tt arXiv:2104.02092}}.

\bibitem{Dillon:2021nxw}
B.~M. Dillon, T.~Plehn, C.~Sauer, and P.~Sorrenson, {\it {Better Latent Spaces
  for Better Autoencoders}},  \href{http://arxiv.org/abs/2104.08291}{{\tt
  arXiv:2104.08291}}.

\bibitem{Finke:2021sdf}
T.~Finke, M.~Kr\"amer, A.~Morandini, A.~M\"uck, and I.~Oleksiyuk, {\it
  {Autoencoders for unsupervised anomaly detection in high energy physics}},
  \href{http://arxiv.org/abs/2104.09051}{{\tt arXiv:2104.09051}}.

\bibitem{Atkinson:2021nlt}
O.~Atkinson, A.~Bhardwaj, C.~Englert, V.~S. Ngairangbam, and M.~Spannowsky,
  {\it {Anomaly detection with Convolutional Graph Neural Networks}},
  \href{http://arxiv.org/abs/2105.07988}{{\tt arXiv:2105.07988}}.

\bibitem{Kahn:2021drv}
A.~Kahn, J.~Gonski, I.~Ochoa, D.~Williams, and G.~Brooijmans, {\it {Anomalous
  Jet Identification via Sequence Modeling}},
  \href{http://arxiv.org/abs/2105.09274}{{\tt arXiv:2105.09274}}.

\bibitem{Aarrestad:2021oeb}
T.~Aarrestad et~al., {\it {The Dark Machines Anomaly Score Challenge: Benchmark
  Data and Model Independent Event Classification for the Large Hadron
  Collider}},  \href{http://arxiv.org/abs/2105.14027}{{\tt arXiv:2105.14027}}.

\bibitem{Dorigo:2021iyy}
T.~Dorigo, M.~Fumanelli, C.~Maccani, M.~Mojsovska, G.~C. Strong, and B.~Scarpa,
  {\it {RanBox: Anomaly Detection in the Copula Space}},
  \href{http://arxiv.org/abs/2106.05747}{{\tt arXiv:2106.05747}}.

\bibitem{Caron:2021wmq}
S.~Caron, L.~Hendriks, and R.~Verheyen, {\it {Rare and Different: Anomaly
  Scores from a combination of likelihood and out-of-distribution models to
  detect new physics at the LHC}},  \href{http://arxiv.org/abs/2106.10164}{{\tt
  arXiv:2106.10164}}.

\bibitem{Govorkova:2021hqu}
E.~Govorkova, E.~Puljak, T.~Aarrestad, M.~Pierini, K.~A. Wo\'zniak, and
  J.~Ngadiuba, {\it {LHC physics dataset for unsupervised New Physics detection
  at 40 MHz}},  \href{http://arxiv.org/abs/2107.02157}{{\tt arXiv:2107.02157}}.

\bibitem{Kasieczka:2021tew}
G.~Kasieczka, B.~Nachman, and D.~Shih, {\it {New Methods and Datasets for Group
  Anomaly Detection From Fundamental Physics}},  7, 2021.
\newblock \href{http://arxiv.org/abs/2107.02821}{{\tt arXiv:2107.02821}}.

\bibitem{Feickert:2021ajf}
M.~Feickert and B.~Nachman, {\it {A Living Review of Machine Learning for
  Particle Physics}},  \href{http://arxiv.org/abs/2102.02770}{{\tt
  arXiv:2102.02770}}.

\bibitem{Shih:2021kbt}
D.~Shih, M.~R. Buckley, L.~Necib, and J.~Tamanas, {\it {Via Machinae: Searching
  for Stellar Streams using Unsupervised Machine Learning}},
  \href{http://arxiv.org/abs/2104.12789}{{\tt arXiv:2104.12789}}.

\bibitem{FCC:2018byv}
{\bf FCC} Collaboration, A.~Abada et~al., {\it {FCC Physics Opportunities}:
  {Future Circular Collider Conceptual Design Report Volume 1}},  {\em Eur.
  Phys. J. C} {\bf 79} (2019), no.~6 474.

\bibitem{FCC:2018evy}
{\bf FCC} Collaboration, A.~Abada et~al., {\it {FCC-ee: The Lepton Collider}:
  {Future Circular Collider Conceptual Design Report Volume 2}},  {\em Eur.
  Phys. J. ST} {\bf 228} (2019), no.~2 261--623.

\bibitem{FCC:2018vvp}
{\bf FCC} Collaboration, A.~Abada et~al., {\it {FCC-hh: The Hadron Collider}:
  {Future Circular Collider Conceptual Design Report Volume 3}},  {\em Eur.
  Phys. J. ST} {\bf 228} (2019), no.~4 755--1107.

\bibitem{Behnke:2013xla}
{\it {The International Linear Collider Technical Design Report - Volume 1:
  Executive Summary}},  \href{http://arxiv.org/abs/1306.6327}{{\tt
  arXiv:1306.6327}}.

\bibitem{Baer:2013cma}
{\it {The International Linear Collider Technical Design Report - Volume 2:
  Physics}},  \href{http://arxiv.org/abs/1306.6352}{{\tt arXiv:1306.6352}}.

\bibitem{Adolphsen:2013jya}
{\it {The International Linear Collider Technical Design Report - Volume 3.I:
  Accelerator \textbackslash{}\& in the Technical Design Phase}},
  \href{http://arxiv.org/abs/1306.6353}{{\tt arXiv:1306.6353}}.

\bibitem{Adolphsen:2013kya}
{\it {The International Linear Collider Technical Design Report - Volume 3.II:
  Accelerator Baseline Design}},  \href{http://arxiv.org/abs/1306.6328}{{\tt
  arXiv:1306.6328}}.

\bibitem{Behnke:2013lya}
H.~Abramowicz et~al., {\it {The International Linear Collider Technical Design
  Report - Volume 4: Detectors}},  \href{http://arxiv.org/abs/1306.6329}{{\tt
  arXiv:1306.6329}}.

\bibitem{InternationalLinearColliderInternationalDevelopmentTeam:2021guz}
{\bf International Linear Collider International Development Team}
  Collaboration, {\it {Proposal for the ILC Preparatory Laboratory (Pre-lab)}},
   \href{http://arxiv.org/abs/2106.00602}{{\tt arXiv:2106.00602}}.

\bibitem{CEPCStudyGroup:2018rmc}
{\bf CEPC Study Group} Collaboration, {\it {CEPC Conceptual Design Report:
  Volume 1 - Accelerator}},  \href{http://arxiv.org/abs/1809.00285}{{\tt
  arXiv:1809.00285}}.

\bibitem{CEPCStudyGroup:2018ghi}
{\bf CEPC Study Group} Collaboration, M.~Dong et~al., {\it {CEPC Conceptual
  Design Report: Volume 2 - Physics \& Detector}},
  \href{http://arxiv.org/abs/1811.10545}{{\tt arXiv:1811.10545}}.

\bibitem{Linssen:1425915}
L.~Linssen, A.~Miyamoto, M.~Stanitzki, and H.~Weerts, {\em {Physics and
  Detectors at CLIC: CLIC Conceptual Design Report}}.
\newblock CERN Yellow Reports: Monographs. CERN, Geneva, 2012.
\newblock Comments: 257 p, published as CERN Yellow Report CERN-2012-003.

\bibitem{CLICdp:2018cto}
{\bf CLICdp, CLIC} Collaboration, T.~K. Charles et~al., {\it {The Compact
  Linear Collider (CLIC) - 2018 Summary Report}},
  \href{http://arxiv.org/abs/1812.06018}{{\tt arXiv:1812.06018}}.

\bibitem{Komiske:2018oaa}
P.~T. Komiske, E.~M. Metodiev, B.~Nachman, and M.~D. Schwartz, {\it {Learning
  to classify from impure samples with high-dimensional data}},  {\em Phys.
  Rev. D} {\bf 98} (2018), no.~1 011502,
  [\href{http://arxiv.org/abs/1801.10158}{{\tt arXiv:1801.10158}}].

\bibitem{Denig:2006kj}
A.~Denig, {\it {The Radiative return: A Review of experimental results}},  {\em
  Nucl. Phys. B Proc. Suppl.} {\bf 162} (2006) 81--89,
  [\href{http://arxiv.org/abs/hep-ex/0611024}{{\tt hep-ex/0611024}}].

\bibitem{Kluge:2008fb}
W.~Kluge, {\it {Initial State Radiation: A Success story}},  {\em Nucl. Phys. B
  Proc. Suppl.} {\bf 181-182} (2008) 280--285,
  [\href{http://arxiv.org/abs/0805.4708}{{\tt arXiv:0805.4708}}].

\bibitem{Czyz:2010hj}
H.~Czyz, A.~Grzelinska, and J.~H. Kuhn, {\it {Narrow resonances studies with
  the radiative return method}},  {\em Phys. Rev. D} {\bf 81} (2010) 094014,
  [\href{http://arxiv.org/abs/1002.0279}{{\tt arXiv:1002.0279}}].

\bibitem{Druzhinin:2011qd}
V.~P. Druzhinin, S.~I. Eidelman, S.~I. Serednyakov, and E.~P. Solodov, {\it
  {Hadron Production via e+e- Collisions with Initial State Radiation}},  {\em
  Rev. Mod. Phys.} {\bf 83} (2011) 1545,
  [\href{http://arxiv.org/abs/1105.4975}{{\tt arXiv:1105.4975}}].

\bibitem{Karliner:2015tga}
M.~Karliner, M.~Low, J.~L. Rosner, and L.-T. Wang, {\it {Radiative return
  capabilities of a high-energy, high-luminosity $e^+e^-$ collider}},  {\em
  Phys. Rev. D} {\bf 92} (2015), no.~3 035010,
  [\href{http://arxiv.org/abs/1503.07209}{{\tt arXiv:1503.07209}}].

\bibitem{Li:2019ufu}
G.~Li, Z.~Li, Y.~Wang, and Y.~Wang, {\it {Improving the measurement of the
  Higgs boson-gluon coupling using convolutional neural networks at $e^+e^-$
  colliders}},  {\em Phys. Rev. D} {\bf 100} (2019), no.~11 116013,
  [\href{http://arxiv.org/abs/1901.09391}{{\tt arXiv:1901.09391}}].

\bibitem{Li:2020vav}
L.~Li, Y.-Y. Li, T.~Liu, and S.-J. Xu, {\it {Learning physics at future $e^-
  e^+$ colliders with machine}},  {\em JHEP} {\bf 10} (2020) 018,
  [\href{http://arxiv.org/abs/2004.15013}{{\tt arXiv:2004.15013}}].

\bibitem{10.5555/3294996.3295098}
M.~Zaheer, S.~Kottur, S.~Ravanbhakhsh, B.~P\'{o}czos, R.~Salakhutdinov, and
  A.~J. Smola, {\it Deep sets},  in {\em Proceedings of the 31st International
  Conference on Neural Information Processing Systems}, NIPS'17, (Red Hook, NY,
  USA), p.~3394–3404, Curran Associates Inc., 2017.

\bibitem{Komiske:2018cqr}
P.~T. Komiske, E.~M. Metodiev, and J.~Thaler, {\it {Energy Flow Networks: Deep
  Sets for Particle Jets}},  {\em JHEP} {\bf 01} (2019) 121,
  [\href{http://arxiv.org/abs/1810.05165}{{\tt arXiv:1810.05165}}].

\bibitem{Alwall:2014hca}
J.~Alwall, R.~Frederix, S.~Frixione, V.~Hirschi, F.~Maltoni, O.~Mattelaer,
  H.~S. Shao, T.~Stelzer, P.~Torrielli, and M.~Zaro, {\it {The automated
  computation of tree-level and next-to-leading order differential cross
  sections, and their matching to parton shower simulations}},  {\em JHEP} {\bf
  07} (2014) 079, [\href{http://arxiv.org/abs/1405.0301}{{\tt
  arXiv:1405.0301}}].

\bibitem{Sjostrand:2014zea}
T.~Sj\"ostrand, S.~Ask, J.~R. Christiansen, R.~Corke, N.~Desai, P.~Ilten,
  S.~Mrenna, S.~Prestel, C.~O. Rasmussen, and P.~Z. Skands, {\it {An
  introduction to PYTHIA 8.2}},  {\em Comput. Phys. Commun.} {\bf 191} (2015)
  159--177, [\href{http://arxiv.org/abs/1410.3012}{{\tt arXiv:1410.3012}}].

\bibitem{Selvaggi:2014mya}
M.~Selvaggi, {\it {DELPHES 3: A modular framework for fast-simulation of
  generic collider experiments}},  {\em J. Phys. Conf. Ser.} {\bf 523} (2014)
  012033.

\bibitem{ILDConceptGroup:2020sfq}
{\bf ILD Concept Group} Collaboration, H.~Abramowicz et~al., {\it
  {International Large Detector: Interim Design Report}},
  \href{http://arxiv.org/abs/2003.01116}{{\tt arXiv:2003.01116}}.

\bibitem{antikt}
M.~Cacciari, G.~P. Salam, and G.~Soyez, {\it The anti-kt jet clustering
  algorithm},  {\em Journal of High Energy Physics} {\bf 2008} (Apr, 2008)
  063--063.

\bibitem{Cacciari:2005hq}
M.~Cacciari and G.~P. Salam, {\it {Dispelling the $N^{3}$ myth for the $k_t$
  jet-finder}},  {\em Phys. Lett. B} {\bf 641} (2006) 57--61,
  [\href{http://arxiv.org/abs/hep-ph/0512210}{{\tt hep-ph/0512210}}].

\bibitem{fastjet}
M.~Cacciari, G.~P. Salam, and G.~Soyez, {\it Fastjet user manual},  {\em The
  European Physical Journal C} {\bf 72} (Mar, 2012).

\bibitem{Barklow:2015tja}
T.~Barklow, J.~Brau, K.~Fujii, J.~Gao, J.~List, N.~Walker, and K.~Yokoya, {\it
  {ILC Operating Scenarios}},  \href{http://arxiv.org/abs/1506.07830}{{\tt
  arXiv:1506.07830}}.

\bibitem{Metodiev:2017vrx}
E.~M. Metodiev, B.~Nachman, and J.~Thaler, {\it {Classification without labels:
  Learning from mixed samples in high energy physics}},  {\em JHEP} {\bf 10}
  (2017) 174, [\href{http://arxiv.org/abs/1708.02949}{{\tt arXiv:1708.02949}}].

\bibitem{Thaler:2010tr}
J.~Thaler and K.~Van~Tilburg, {\it {Identifying Boosted Objects with
  N-subjettiness}},  {\em JHEP} {\bf 03} (2011) 015,
  [\href{http://arxiv.org/abs/1011.2268}{{\tt arXiv:1011.2268}}].

\bibitem{Thaler:2011gf}
J.~Thaler and K.~Van~Tilburg, {\it {Maximizing Boosted Top Identification by
  Minimizing N-subjettiness}},  {\em JHEP} {\bf 02} (2012) 093,
  [\href{http://arxiv.org/abs/1108.2701}{{\tt arXiv:1108.2701}}].

\bibitem{chollet2015keras}
F.~Chollet et~al., {\it Keras},  2015.
\newblock https://github.com/fchollet/keras.

\bibitem{tensorflow2015-whitepaper}
M.~A. et~al, {\it {TensorFlow}: Large-scale machine learning on heterogeneous
  systems},  2015.
\newblock Software available from tensorflow.org.

\bibitem{adam}
D.~P. Kingma and J.~Ba, {\it Adam: A method for stochastic optimization},
  \href{http://arxiv.org/abs/1412.6980}{{\tt arXiv:1412.6980}}.

\bibitem{adagrad}
J.~Duchi, E.~Hazan, and Y.~Singer, {\it Adaptive subgradient methods for online
  learning and stochastic optimization},  {\em J. Mach. Learn. Res.} {\bf 12}
  (July, 2011) 2121–2159.

\bibitem{rmsprop}
G.~Hinton, N.~Srivastava, and K.~Swersky, ``Neural networks for machine
  learning: Lecture 6a.''
  \url{http://www.cs.toronto.edu/~tijmen/csc321/slides/lecture_slides_lec6.pdf}.

\bibitem{lny23_1}
A.~Banfi, G.~P. Salam, and G.~Zanderighi, {\it {Phenomenology of event shapes
  at hadron colliders}},  {\em JHEP} {\bf 06} (2010) 038,
  [\href{http://arxiv.org/abs/1001.4082}{{\tt arXiv:1001.4082}}].

\bibitem{lny23_2}
R.~B. et~al, {\it Studies of quantum chromodynamics with the {{ALEPH}}
  detector},  {\em Physics Reports} {\bf 294} (1998), no.~1 1--165.

\bibitem{eventshape_opal}
{\bf OPAL Collaboration} Collaboration, {\it {QCD studies with e+e-
  annihilation data at 130 and 136 GeV}},  {\em Z. Phys. C} {\bf 72} (Mar,
  1996) 191--206. 41 p.

\bibitem{eventshape_atlas}
G.~Aad, B.~Abbott, J.~Abdallah, S.~Abdel~Khalek, A.~A. Abdelalim, O.~Abdinov,
  B.~Abi, M.~Abolins, O.~S. AbouZeid, and et~al., {\it {Measurement of event
  shapes at large momentum transfer with the ATLAS detector in pp collisions at
  $\sqrt{\mathbf{s} }=7\ \mathrm{TeV}$}},  {\em The European Physical Journal
  C} {\bf 72} (Nov, 2012).

\bibitem{totaljetmass}
A.~M. Sirunyan, A.~Tumasyan, W.~Adam, F.~Ambrogi, E.~Asilar, T.~Bergauer,
  J.~Brandstetter, M.~Dragicevic, J.~Erö, and et~al., {\it Event shape
  variables measured using multijet final states in proton-proton collisions at
  $\sqrt{s}=13$ tev},  {\em Journal of High Energy Physics} {\bf 2018} (Dec,
  2018).

\end{thebibliography}\endgroup


\providecommand{\href}[2]{#2}\begingroup\raggedright\endgroup
